\documentclass[letterpaper,11pt]{article}
\pdfoutput=1
\usepackage{jheppub}

\allowdisplaybreaks

\usepackage{gensymb}
\usepackage{braket}
\usepackage{amsmath,amssymb,amsthm,amsfonts}
\usepackage{graphicx}
\usepackage{subfigure}
\usepackage{dcolumn}	 
\usepackage{hyperref}   
\usepackage{bm}		 
\usepackage{epsfig}
\usepackage{epstopdf}
\usepackage{setspace}
\usepackage{tabularx}
\usepackage[usenames,dvipsnames]{xcolor}
\usepackage{slashed}
\usepackage{comment}
\usepackage{enumitem}
\usepackage{tikz}

\usepackage[most]{tcolorbox}

\tcbset{colback=yellow!10!white, colframe=red!50!black, 
        highlight math style= {enhanced, %<-- needed for the ’remember’ options
            colframe=red,colback=red!10!white,boxsep=0pt}
        }

\newcommand{\be}{\begin{equation}\begin{aligned}}
\newcommand{\ee}{\end{aligned}\end{equation}}

\definecolor{green2}{rgb}{0,0.56,0.32}

\RequirePackage[normalem]{ulem}

\newcommand{\m}{\text{m}}
\newcommand{\mev}{\text{~MeV}}
\newcommand{\gev}{\text{~GeV}}

\newcommand{\tanb}{\tan \beta}
\def\cosba{\cos(\beta-\alpha)}
\def\mC{m_{H^\pm}}

\def\figureautorefname~#1\null{Fig.\,#1\null}
\def\tableautorefname~#1\null{Tab.\,#1\null}

\def\equationautorefname~#1\null{Eq.\,(#1)\null}
\def\gev{\textrm{~GeV}}

\begin{document}

\title{Light Scalars at FASER}
%\author{Felix Kling$^{a}$, }
 \author{Shuailong Li$^{\flat}$,}
 \author{Huayang Song$^{\natural}$,}
 \author{Shufang Su$^{\flat}$,}
 \author{Wei Su$^{\dagger}$}
 \affiliation{
$^{\dagger}$School of Science, Shenzhen Campus of Sun Yat-sen University, No. 66, Gongchang Road, \\ Guangming District, Shenzhen, Guangdong 518107, P.R. China \\
$^{\flat}$Department of Physics, University of Arizona, Tucson, AZ  85721, USA\\
$^{\natural}$CAS Key Laboratory of Theoretical Physics, Institute of Theoretical Physics, Chinese Academy of Sciences, Beijing 100190, China\\}

\emailAdd{shuailongli@email.arizona.edu}
\emailAdd{huayangs@itp.ac.cn}
\emailAdd{shufang@email.arizona.edu}
\emailAdd{suwei26@mail.sysu.edu.cn}

% \begin{abstract}
\abstract{FASER, the ForwArd Search ExpeRiment,  is a currently operating experiment at the Large Hadron Collider (LHC) that can detect light long-lived particles produced in the forward region of the LHC interacting point.  In this paper, we study the prospect of detecting light CP-even and CP-odd scalars at FASER and FASER 2.  Considering a model-independent framework describing the most general interactions between a CP-even or CP-odd scalar and SM particles using the notation of
coupling modifiers in the effective Lagrangian, we develop the general formalism for the scalar production and decay.   We then analyze the FASER and FASER 2 reaches of light scalars in the large $\tan\beta$ region of the Type-I two Higgs double model as a case study, in which light scalars with relatively long lifetime could be accommodated.  In the two benchmark scenarios we considered, the light (pseudo)scalar decay length varies in $(10^{-8}, 10^5)$ meters.  Both FASER and FASER 2 can probe a large part of the parameter space in the large $\tan\beta$ region up to $10^5$, extending beyond the constraints of the other existing experiments.    }
% \end{abstract}
%\titlepage
\maketitle
% \tableofcontents

\newpage
\section{Introduction} 
Searches for beyond the Standard Model (SM) new particles are one of main physics drivers at the Large Hadron Collider (LHC) after the discovery of the SM-like Higgs. Traditional resonance searches at the LHC main detectors, like ATLAS and CMS, are aimed at promptly decaying particles with electroweak-scale masses and $\mathcal{O}(1)$ couplings to the SM particles. Recently there are increasing  interests in the detection of long-lived particles (LLPs) at the LHC main detectors ATLAS and CMS~\cite{Gershtein:2017tsv, Liu:2018wte, Lee:2018pag, Alimena:2019zri, Liu:2019ayx, Liu:2020vur, Gershtein:2020mwi, Chiu:2021sgs, Fischer:2021sqw, Bose:2022obr, ATLAS:2022pib, ATLAS:2022cob, CMS:2016ybj, CMS:2015pca, CMS:2022qej, ATLAS:2018rjc, CMS:2017kku, ATLAS:2022zhj, CMS:2020iwv, ATLAS:2021mdj, CMS:2019zxa, CMS:2019qjk, ATLAS:2019qrr, CMS:2021sch, ATLAS:2022vyq, ATLAS:2022gvp}, MATHUSLA~\cite{Chou:2016lxi, Curtin:2017izq, Evans:2017lvd, Curtin:2018mvb, Curtin:2018ees, Berlin:2018jbm, MATHUSLA:2018bqv,  MATHUSLA:2019qpy, Alimena:2019zri, Jodlowski:2019ycu, Alidra:2020thg, MATHUSLA:2020uve, MATHUSLA:2022sze, Bose:2022obr}, CodexB~\cite{Gligorov:2017nwh, Dey:2019vyo, Aielli:2019ivi, Aielli:2022awh}, ANUBIS~\cite{Bauer:2019vqk, Hirsch:2020klk, Dreiner:2020qbi}, etc.  However, those searches are mostly sensitive to particles produced in the transverse region. For the light  LLPs produced mostly in the forward region, the sensitivity is greatly reduced.
 
Light LLPs with mass  a few GeV or lighter are predicted in many new physics scenarios~\cite{Egana-Ugrinovic:2019wzj, Kitahara:2019lws, Liu:2020qgx, Kling:2020mch, Baum:2021qzx, Athron:2021iuf}, in particular, those involve dark matter~\cite{Tucker-Smith:2001myb, Dienes:2011ja, Dienes:2011sa, Dienes:2012jb, Hochberg:2015vrg}, hidden valley~\cite{Strassler:2006im, Strassler:2006ri}, dark photon~\cite{Holdom:1985ag, Bauer:2018onh, Fabbrichesi:2020wbt, Caputo:2021eaa}, axion-like particles~\cite{Peccei:1977hh, Peccei:1977ur, Jaeckel:2010ni, Bauer:2017ris} or heavy neutral leptons~\cite{Gell-Mann:1979vob, Mohapatra:1979ia, Schechter:1980gr, Asaka:2005pn, Kersten:2007vk, Drewes:2015iva}.  There are several undergoing and proposed experiments for light LLPs searches, including  Belle-II~\cite{Belle-II:2010dht, Belle-II:2018jsg, Filimonova:2019tuy}, FNAL-$\mu$~\cite{Chen:2017awl, Marsicano:2018vin}, HPS~\cite{Celentano:2014wya, HPS:2018xkw, Moreno:2018tlx}, NA62~\cite{Lanfranchi:2017wzl, NA62:2017rwk}, NA64~\cite{NA64:2016oww, Gninenko:2019qiv, Banerjee:2019pds, NA64:2020qwq}, NA64$_e^{++}$~\cite{Gninenko:2300189}, NA64$_\mu$~\cite{Gninenko:2640930}, SeaQuest~\cite{SeaQuest:2017kjt, Liu:2017ryd}, SpinQuest/DarkQuest~\cite{Berlin:2018pwi, Batell:2020vqn, Blinov:2021say, Apyan:2022tsd}, LongQuest~\cite{Tsai:2019buq}, and SHiP~\cite{Bonivento:2013jag, Alekhin:2015byh, SHiP:2015vad}.  Those light particles could also be copiously produced in the forward region of a high energy collider, for example, the LHC.    FASER, the ForwArd Search ExpeRiment, is designed to detect LLPs produced at the ATLAS interaction point (IP),   traveling in the very forward region,  and decaying in FASER (480 meters from the IP) into two very energetic particles~\cite{Feng:2017uoz, FASER:2018ceo, FASER:2018bac, FASER:2022hcn, FASER:2021ljd,FASER:2021cpr}.   FASER has been taking data since summer, 2022.    Given the distinctive signature and low background environment,  FASER provides a unique opportunity to probe light particles with suppressed couplings~\cite{Feng:2017uoz, FASER:2018eoc, FASER:2018bac}, for example, dark photons~\cite{Feng:2017uoz}, axion-like particles~\cite{Feng:2018pew}, and heavy neutral leptons~\cite{Kling:2018wct}.   At the HL-LHC,   FASER will be upgraded to FASER 2 with a larger volume of the detector, potentially at the same location~\cite{FASER:2018eoc} or at the Forward Physics Facility (FPF)~\cite{Anchordoqui:2021ghd, Feng:2022inv}.  In this paper, we explore the collider reach of light long lived scalars at FASER and FASER 2 experiments.

The simplest extension of the SM with a long-lived scalar is the dark Higgs scenario in which a new singlet scalar $S$ mixes with the SM Higgs with the mixing angle $\theta$.   The couplings of the new physical scalar with SM particles follow those of the SM Higgs, 
re-scaled by $\sin\theta$ at leading order. Previous studies of this scenario at FASER show significant sensitivity to a light long-lived   scalar~\cite{Feng:2017vli, Feng:2018pew, FASER:2018eoc, Kling:2021fwx}.  However,  these studies are quite limited given the specific coupling structure of the light scalar with SM particles.

In this work we consider a model-independent framework describing the most general interactions between a CP-even or CP-odd scalar and SM particles using the notation of coupling modifiers in the effective Lagrangian. We develop general formalism for  the productions of the light scalar from meson decays, as well as  re-analyse the scalar decay rates. Given the  non-universal couplings of the scalar to SM particles, the light scalar behaves much more complicatedly compared to the one in the simplest dark Higgs scenario. The CP-odd scalar can further mix with the light meson states, resulting in more non-standard features.    We evaluate the decays of the light scalars to diphotons, dileptons, as well as hadronic final states.    In particular, for mass below about 2 GeV, we adopt chiral perturbation theory  and dispersive analysis to calculate the hadronic decay of   light scalars.   For heavier masses, the spectator model is applied.  We  develop a  program~\cite{LSDedcay} to calculate the decays of a light CP-even or CP-odd scalar across a wide mass range, incorporating the coupling modifiers of the light scalars to the SM particles.   Our program can be applied to various new physics models with a light scalar. 

Sub-GeV scalar arises in various well-studied models, such as the  two Higgs doublet model (2HDM)~\cite{Cherchiglia:2017uwv}, the Next-to-Minimal
2HDM (N2HDM)~\cite{Liu:2020qgx} and the Next-to-Minimal Supersymmetric Standard Model (NMSSM)~\cite{Domingo:2016yih}.  As a case study, we apply our general formalism on the  productions and decays of light scalars in the framework of the 2HDMs. After considering both theoretical and experimental constraints on the 2HDMs, we find that the most viable scenario that could accommodate a light long lived scalar is the large $\tan\beta$ region of the Type-I 2HDM.  We  identify two specific benchmark regions that could accommodate a long-lived light scalar or pseudoscalar.  We  further analyze the reach of FASER and FASER 2 on the parameter space of the Type-I 2HDM for theses two benchmark regions. 

The paper is organized as follows.  In Sec.~\ref{sec:evenscalar} we introduce the general interactions of CP-even scalar and discuss  the production and decays of a light CP-even scalar. We present the study for the light CP-odd scalar in Sec.~\ref{sec:oddscalar}, emphasizing the special features due to its mixing with meson states.  Our case study of the large $\tan\beta$ region of the Type-I 2HDM and main results are presented in Sec.~\ref{sec:2HDM}. We conclude  in Sec.~\ref{sec:conclusion}. Expressions for loop-induced form factors, tri-meson decay amplitudes, tri-Higgs couplings, and effective flavor changing couplings of (pseudo)scalars in the Type-I 2HDM are collected in the Appendix.

% --------------------------------------------------------------------
\section{Light CP-even Scalar}
\label{sec:evenscalar}
 
\subsection{Effective Lagrangian}
The effective Lagrangian for a (light) CP-even scalar $\phi$ interacting with SM particles can be written as~\cite{Zyla:2020zbs}
\begin{eqnarray}
 \mathcal{L} &=& - \frac{1}{2}{m_\phi^{\textcolor{black}{2}}} \phi^2 - \sum_f  {\xi_{\phi}^{f}}  \frac{m_f}{v} \, \phi \bar{f} f    {+} {\xi_{\phi}^{W}} \frac{2m_W^2}{v} \, \phi W^{\mu+} W^-_\mu {+} {\xi_{\phi}^{Z}} \frac{m_Z^2}{v} \, \phi Z^{\mu} Z_\mu  \nonumber \\ 
 &&+ {\xi_{\phi}^{ g}} \frac{\alpha_s}{12\pi v} \phi G_{\mu\nu}^a G^{a\mu\nu}  +  {\xi_{\phi}^{\gamma}} \frac{\alpha_{\rm ew}}{4\pi v} \phi F_{\mu\nu} F^{\mu\nu}, 
\label{eq:phi_lag}
\end{eqnarray}
where $F_{\mu\nu}$ and $G_{\mu\nu}^a$ denote the field-strength tensors for the photon and gluon fields respectively.  Various $\xi$s are the coupling modifiers for the interactions between $\phi$ and SM particles.
 
In the SM, the effective couplings $\phi gg$ and $\phi \gamma\gamma$ are loop generated. The contribution of quarks, leptons and $W$ bosons to $\xi_{\phi}^{g}$ and $\xi_{\phi}^{\gamma}$ are given by \cite{Tanabashi:2018oca,Djouadi:2005gi,Fugel:2004ug} 
\begin{align}
    &\xi^{g}_{\phi}=\sum_{f\in q}\frac{3}{2}\xi^i_\phi \mathcal{A}_{1 / 2}^{\phi}(\tau_f^\phi), 
        \label{eq:coup_hgg}
\\
    &\xi_\phi^\gamma=\sum_{f\in q, \ell} N_{c} Q_{f}^{2} \xi_\phi^{f} \mathcal{A}_{1 / 2}^{\phi}\left(\tau_{f}^\phi\right)+ \xi_\phi^{V} \mathcal{A}_{1}^{\phi}\left(\tau_{W}^\phi\right).  
    \label{eq:coup_hgaga}
\end{align}
 Here $\tau_{i}^\phi=m_{\phi}^{2} / 4 m_{i}^{2}$ and $m_i$ is the mass of the particle running in the loop.  The expressions for the form factors  $\mathcal{A}_{1/2,1}^\phi$ for fermions and gauge bosons can be found in   Appendix~\ref{app:phigg_phigaga}.   
 For new physics models with new charged/colored particles coupling to $\phi$, additional contributions to $\xi^{\gamma}_{\phi}$ and $\xi^{g}_{\phi}$ are possible. 

%*****************************************************
\subsection{Productions}
%*****************************************************
\label{sec:cp-even-pro}

Light scalar $\phi$ is mainly produced in the decay of hadrons~\cite{Boiarska:2019jym, Feng:2017vli,Leutwyler:1989xj,Gunion:1989we,Egana-Ugrinovic:2019wzj,Chivukula:1988gp,Grinstein:1988yu}, the semileptonic decays of pions and kaons~\cite{Dawson:1989kr,Feng:2017vli},  as well as radiative bottomonium decay~\cite{Winkler:2018qyg}. Another production mode of light scalars is through their bremsstrahlung in proton-proton collisions~\cite{Boiarska:2019jym}.   The light scalar can also be produced via $h \to \phi\phi$.  However, the $h\phi\phi$ coupling can not be too large given the invisible Higgs decay constraints.     $Z$ and $W$ decays could also contribute to the production of $\phi$, which typically has a high transverse momentum.   In the forward region of the LHC IP, the contribution to the production of light CP-even scalar $\phi$ from the last three channels are small~\cite{Feng:2017vli, Boiarska:2019jym}. Therefore we do not include them and focus on the meson decay processes instead. 
 
The light scalar $\phi$ can be produced in meson decays via flavor changing effects. The corresponding effective Lagrangian of flavor changing quark interactions with the scalar $\phi$   can be defined as~\cite{Boiarska:2019jym}
\begin{equation}
    \mathcal{L}_{eff}=\frac{\phi}{v}\sum\xi^{ij}_{\phi}m_{f_j}\bar{f}_i P_R f_j+h.c.
    \label{eq:Lfceff_phi}
\end{equation}
where $\xi_{\phi}^{ij}$ are the effective couplings for quarks $f_i$ and $f_j$, and $P_R\equiv(1+\gamma_5)/2$.   
$\xi_\phi^{ij}$ in various beyond the SM  (BSM) scenarios can be obtained   via tree and/or loop level contributions.

\begin{description}
\item[Heavy $B$ Meson Decays]
 The inclusive decay of $B$ mesons into ligth CP-even scalar is dominated by the above flavor changing effective interaction between $b$ and $s$ quarks. Uncertainties from strong interaction effects are minimized in the ratio~\cite{Grinstein:1988yu,Chivukula:1988gp}
\begin{equation}
    \frac{\text{Br}(B\rightarrow X_s\phi)}{\text{Br}(B\rightarrow X_c e\nu)}=\frac{\Gamma(b\rightarrow s\phi)}{\Gamma(b\rightarrow ce\nu)}=\frac{12\pi^2 v^2}{m_b^2}(1-\frac{m_\phi^2}{m_b^2})^2\frac{1}{f(m_c^2/m_b^2)}\left|\frac{\xi^{bs}_\phi}{V_{cb}}\right|^2,
\end{equation}
where $X_{s,c}$  denotes any strange and charm hadronic state, and $f(x)=(1-8x+x^2)(1-x^2)-12x^2\ln{x}$ is the phase space factor. We take  $\text{Br}(B\rightarrow X_c e\nu)=0.104$ for both $B^0$ and $B^\pm$ from Ref.~\cite{Zyla:2020zbs}.   

\begin{figure}[htb]
\begin{center}
\includegraphics[width=0.4\textwidth]{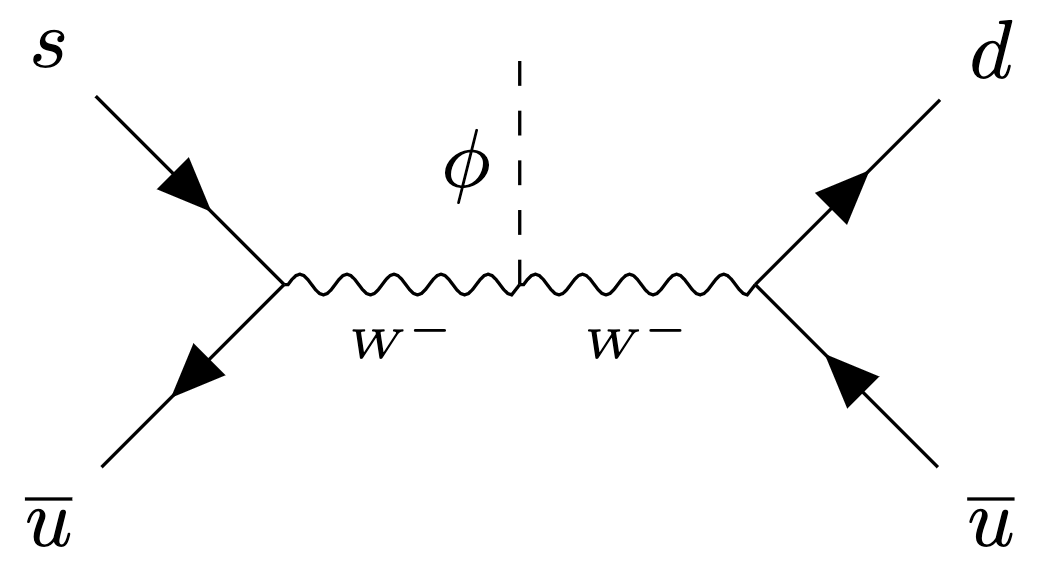}
\end{center}
\caption{SM contribute to the transition $K^-\rightarrow \pi^-\phi$ via effective four quark operator.     \label{Fig:4Quark}}
\end{figure}

\item[Kaon Decays]
In addition to the flavor changing quark interactions mentioned above, four quark operators can also contribute non-negligibly to the two-body kaon decays. The corresponding Feynman diagram with SM contribution for $K^-\rightarrow \pi^-\phi$ is shown in Fig.~\ref{Fig:4Quark}, 
which results in an effective four-fermion-scalar interaction
\begin{align}
    \mathcal{L}_{eff}=&\frac{2^3 G_F^{3/2}}{2^{1/4}}\xi_{\phi}^W V_{ud}^* V_{us}\bar{d}\gamma^\mu P_L u\bar{u}\gamma_\mu P_L s\phi +h.c.  
     \label{Eq:Lag_4FH}
\end{align}
Including both contributions,   the total amplitude for $K^\pm \rightarrow \pi^\pm\phi$ is~\cite{Leutwyler:1989xj, Gunion:1989we, Bezrukov:2009yw, Feng:2017vli, Boiarska:2019jym} 
\begin{eqnarray}
    \mathcal{M} (K^\pm \rightarrow \pi^\pm\phi) &=&G_F^{1/2}2^{1/4}\xi_{\phi}^W\left[\frac{7\lambda(m_{K^\pm}^2+m_{\pi^\pm}^2-m_\phi^2)}{18}-\frac{7A_{K^\pm}m_{K^\pm}^2}{9}\right] \nonumber \\ &+&\frac{\xi^{ds}_\phi}{2v}m_{s}\frac{m_{K^\pm}^2-m_{\pi^\pm}^2}{m_{s}-m_{d}}f_0^{K^\pm \pi^\pm}(q^2),
    \label{Eq:Matrix_4FH}
\end{eqnarray}
  where $\lambda\simeq 3.1\times 10^{-7}$, $A_{K^\pm} \approx 0$~\cite{Leutwyler:1989xj, Gunion:1989we, Bezrukov:2009yw, Feng:2017vli}, and $f_0^{K^\pm\pi^\pm}(q^2)$ is the form-factor taken to be $0.96$~\cite{Boiarska:2019jym}.     The corresponding branching fraction is
\begin{equation}
    \text{Br}(K^\pm\rightarrow \pi^\pm \phi)=\frac{1}{\Gamma_{K^\pm}}\frac{2p_\phi^0}{m_{K^\pm}}\frac{|\mathcal{M}|^2}{16\pi m_{K^\pm}}  ,
 \end{equation}
where $p_\phi^0$ is the magnitude of the $\phi$  momentum in the parent meson’s rest frame.     Expressions for the neutral $K_L$ and $K_S$ decay can be obtained similarly~\cite{Leutwyler:1989xj, Gunion:1989we, Bezrukov:2009yw, Feng:2017vli}.     

\item[$\eta^{(\prime)}$ Decays]
CP-even scalar can also be produced in the decays of $\eta$ and $\eta'$. The branching fraction of $\eta^{(\prime)}$ meson to a scalar and pion is given by
\begin{equation}
    \text{Br}(\eta^{(\prime)}\rightarrow\pi\phi)=\frac{1}{\Gamma_{\eta^{(\prime)}}}\frac{2p_\phi^0}{m_{\eta^{(\prime)}}}\frac{|g_{\phi\eta^{(\prime)}\pi}|^2}{16\pi m_{\eta^{(\prime)}}}.
\end{equation}   
The coupling $g_{\phi\eta^{(\prime)}\pi}$ can be obtained using chiral perturbation theory as~\cite{Batell:2018fqo, Egana-Ugrinovic:2019wzj}
\begin{equation}
    g_{\phi\eta^{(\prime)}\pi}=-\frac{1}{v}\left[m_u\xi_\phi^u-m_d\xi_\phi^d+\frac{2}{9}\left(m_u-m_d\right)\left(\xi_\phi^g+\sum_{q=c, b, t}\xi_\phi^q\right)\right]c_{\phi\eta^{(\prime)}\pi}\tilde{B}.
\end{equation}
where $\tilde{B}=m_\pi^2/(m_u+m_d)\simeq 2.6~\gev$, and $c_{\phi\eta^{(\prime)}\pi}=(\cos\theta_\eta\pm\sqrt{2}\sin\theta_\eta)/\sqrt{3}$.    The mixing angle $\theta_\eta$ between $\eta$ and $\eta'$ can be obtained from experiments, which is taken to be $-13^\circ$~\cite{Domingo:2016yih}.    
 
\item[Semileptonic Decays of Mesons] 
Besides the two-body hadronic decays of mesons discussed above, the 3-body semileptonic decays of mesons can also produce light scalars. The branching fraction for $X\rightarrow\phi e\nu$ is~\footnote{We do not consider the process $X\rightarrow\phi \mu\nu$ due to the reduced phase space for such process for $K, \pi$ decays,  which significantly suppresses the corresponding branching fraction.} ~\cite{Chivukula:1988gp, Dawson:1989kr, Cheng:1989ib, Boiarska:2019jym}
 \begin{equation}
    \text{Br}(X\rightarrow\phi e\nu)=\frac{\sqrt{2}G_F m_X^4 {|{\xi_\phi^W}|^2}}{96\pi^2 m_\mu^2(1-m_\mu^2/m_X^2)^2}\times\text{BR}(X\rightarrow\mu\nu)f(\frac{m_\phi^2}{m_X^2})\left(1-\frac{2n_h}{33-2n_l}\right)^2,
\end{equation}
where $f(x)$ is the phase space factor motioned previously, and $n_h$ and $n_l$ are numbers of heavy and light quarks in the corresponding EFT describing the meson $X$, respectively.   For leptonic decays of light mesons pion and kaon,  $n_h=n_l=3$.  
 
\item[Radiative Bottomonium $\Upsilon$ Decay]
A light scalar can be produced in the radiative decay of bottomonium $\Upsilon\rightarrow\gamma\phi$. It is convenient to express the corresponding branching ratio in the form~\cite{Winkler:2018qyg}  
\begin{equation}
    \frac{\text{Br}(\Upsilon\rightarrow\gamma\phi)}{\text{Br}(\Upsilon\rightarrow e^+e^-)}=\frac{ G_F m_b^2|{\xi_\phi^b}|^2}{\sqrt{2}\pi\alpha}(1-\frac{m_\phi^2}{m_\Upsilon^2})\times \frac{2}{3}  \left(1-\frac{m_\phi^6}{m_\Upsilon^6}\right),
\end{equation}
where the last term is a fitted correction function which reproduces the NLO corrections described in  Ref.~\cite{Ellis:1985yb}. 

  \end{description}
  
%*****************************************************
\subsection{Decays of Light CP-even Scalar}
%*****************************************************
\label{sec:Hdecay}

 Depending on the mass of the CP-even scalar $\phi$, it can decay into pair of photons, leptons, and multiple hadrons or pair of quarks.  For $m_\phi \lesssim 2$ GeV,  a dispersive analysis method introduced in Ref.~\cite{Winkler:2018qyg} is used to calculate the partial decay width into hadrons, while for $m_\phi \gtrsim 2$ GeV, the perturbative spectator model is applied. 

\begin{description}
\item[Decays into Diphoton]

 The decay rate of a CP-even scalar into diphoton is given by 
\begin{equation}
\Gamma_{ \gamma \gamma}=\frac{G_{F} \alpha_{\rm ew}^{2} m_{\phi}^{3}}{32 \sqrt{2} \pi^{3}}  \Big| \xi_\phi^\gamma
\Big|^{2}.
\label{eq:hgaga}
\end{equation}

\item[Decays into Leptons]
The decay rate of a CP-even scalar into leptonic final states can be calculated using perturbation theory.  At leading order, the partial decay width is~\cite{Winkler:2018qyg} 
\be
    \Gamma_{\ell^+ \ell^-} =\frac{G_F m_\phi m_\ell^2\beta_\ell^3 }{4\sqrt{2}\pi}|\xi_\phi^{\ell}|^2,
\ee
with $\ell=e, \mu, \tau$. Here $\beta_\ell=\sqrt{1-4m_\ell^2/m_\phi^2}$ is the velocity of the leptons in the rest frame of $\phi$.  
%%%%%%%%
\item[Hadronic Decays into Pions and Kaons for $\mathbf{m_\phi \lesssim 2}~\gev$]
For $ m_\phi\lesssim 2~\gev$, we have to use a hadronic picture since quarks cannot be treated as free particles and partonic picture fails.   Given the parton level Lagrangian of Eq.~(\ref{eq:phi_lag}), the decay rate of pion and kaon pairs for a light CP-even scalar $\phi$ is~\cite{Donoghue:1990xh}
\begin{eqnarray}
&&\Gamma_{\pi \pi}=\frac{3 G_{F}}{16 \sqrt{2} \pi m_\phi} \beta_{\pi}\left|
\xi^{g}_{\phi}\frac{2}{27}(\Theta_{\pi}-\Gamma_{\pi}-\Delta_{\pi})+
\frac{m_u \xi_{\phi}^{u}+m_d \xi_{\phi}^{d}}{m_u+m_d} \Gamma_{\pi} + (\xi_{\phi}^{s})\Delta_{\pi}
\right|^{2}, \\
&&\Gamma_{K K}=\frac{ G_{F}}{4 \sqrt{2} \pi m_\phi} \beta_{K}\left|
\xi^{g}_{\phi}\frac{2}{27}(\Theta_{K}-\Gamma_{K}-\Delta_{K})+
\frac{m_u \xi_{\phi}^{u}+m_d \xi_{\phi}^{d}}{m_u+m_d} \Gamma_{K} + (\xi_{\phi}^{s})\Delta_{K}
\right|^{2},
\label{eq:width}
\end{eqnarray}
with with $\beta_{i}=\sqrt{1-4 m_{i}^{2} / m_{\phi}^{2}}$.   
$\Theta_{\pi,K}$, $\Gamma_{\pi,K}$ and $\Delta_{\pi,K}$ are form factors that need to be evaluated at $\sqrt{s}=m_{\phi}$.  In the chiral perturbation theory estimation, higher orders are suppressed by powers of the chiral symmetry breaking scale $\Lambda_{\chi} \sim 1~\gev$, which could be sizable for $m_{\phi} \gtrsim 0.5~\gev$.   In our analyses, we use the form factors extracted through dispersion relations~\cite{Winkler:2018qyg, Monin:2018lee} to take into account the higher order effects.

\item[Further Hadronic Decays for $\mathbf{m_\phi\lesssim 2}~\gev$]
At larger masses, $m_\phi>m_{4\pi}$, additional decay channels into further hadronic final states open up. These include the decays $\phi \to 4\pi, \eta\eta, KK \pi\pi, \rho\rho \ldots$, with decay width of~\cite{Winkler:2018qyg, Grinstein:1988yu},
\begin{equation}
    \Gamma_{4 \pi, \eta \eta, \rho \rho, \ldots}=C |\xi_\phi^{g}|^2 m_{\phi}^{3} \beta_{2\pi}.
\end{equation}
$C$ is set to $5.1\times 10^{-9}~\gev^{-2}$ to 
obtain smooth hadronic decay rate transiting into the rate of the spectator model at $m_\phi=2~\gev$.  

\item[Decays into Quarks for $\mathbf{m_\phi\gtrsim 2}~\gev$]
The perturbative spectator model can be applied for hadronic decays for higher scalar masses.  The ratios of the decay rates to quarks comparing to that to dilepton are given by~\cite{Winkler:2018qyg}, 
\be
    \Gamma_{\ell^+\ell^-}:\Gamma_{s\bar{s}}:\Gamma_{c\bar{c}}:\Gamma_{b\bar{b}}=
    |\xi_\phi^{\ell}|^2  m_\ell^2\beta_\ell^3:3|\xi_\phi^s|^2 m_s^2\beta_K^3:3 |\xi_\phi^c|^2 m_c^2\beta_D^3:3 |\xi_\phi^b|^2 m_b^2\beta_B^3,
\ee
in which we set $m_s=95~\mev$, $m_c=1.3~\gev$ and $m_b=4.18~\gev$. The kinematic threshold is set by the lightest meson containing an $s$, $c$, or $b$ quark respectively: $m_K=493.677~\mev$ ($K^\pm$), $m_D=1864.84~\mev$ ($D^0$ meson) and $m_B=5279.15~\mev$ ($B^\pm$).

\item[Decays into Gluons for $\mathbf{m_\phi\gtrsim 2}~\gev$]

We also consider loop induced decays into gluon pairs. The corresponding decay width is given by
\begin{equation}
    \Gamma_{gg}
    =\frac{G_F \alpha_s^2  m_\phi^3 }{36\sqrt{2}\pi^3}|\xi_{\phi}^{g}|^2,
\label{eq:hgg}
\end{equation}
with $\alpha_s(m_\phi)$  taken from Ref.~\cite{Bethke:2006ac}.

\end{description}

\section{Light CP-odd Scalar}
\label{sec:oddscalar}
\subsection{Effective Lagrangian}
The effective Lagrangian involving CP-odd scalar $A$ and its interaction with SM particles can be expressed as~\footnote{Note that unlike $\xi_\phi^g$ defined in Eq.~(\ref{eq:phi_lag}), $\xi_A^g$ is usually defined with $\frac{\alpha_s}{4\pi v }$ instead of $\frac{\alpha_s}{12\pi v}$ factored out.}~\cite{Domingo:2016yih}  
\begin{equation}
    \mathcal{L}_{A}=-\frac{1}{2}m_A^2A^2+ \sum_{f=u, d, e} \xi_{A}^f \frac{im_f}{v}  \bar{f} \gamma_{5} f A+ \xi_A^g \frac{\alpha_s}{4\pi v} A G_{\mu\nu}^a\tilde G^{a \mu\nu}+ \xi_A^\gamma \frac{\alpha_{\rm ew}}{4\pi v} A F_{\mu\nu}\tilde F^{\mu\nu},
    \label{eq:Aeff}
\end{equation}
where $\tilde F_{\mu\nu}\equiv 1/2 \varepsilon^{\mu\nu\rho\sigma}F_{\rho\sigma}$ for completely anti-symmetric symbol $\varepsilon^{\mu\nu\rho\sigma}$, and $\tilde G$ is defined similarly.     
 The SM contributions to loop-induced effective couplings, $\xi_A^\gamma$ and $\xi_A^g$, are given by~\cite{Domingo:2016yih,Djouadi:2005gj}
 \begin{align}
    &\xi_A^g=-\frac{1}{4}\sum_{f \in q}\xi_{A}^f\mathcal{A}_{1/2}^A(\tau_f^A),
    \label{eq:xiAgluon}\\
    &\xi_A^\gamma=-\frac{1}{2} \sum_{f \in q,\ell}N_c^f Q_f^2\xi_{A}^f\mathcal{A}_{1/2}^A(\tau_f^A).
\label{eq:xiAgamma}
\end{align}
The expression for $\mathcal{A}_{1/2}^A$ can be found in Appendix~\ref{app:phigg_phigaga}.
  
The pseudoscalar $A$ shares its quantum
numbers with some of the mesons (e.g. $\pi^0$, $\eta$ and $\eta^\prime$), which typically induce mixings among these states. We will still use the notation $A$ to refer to the mass eigenstate which contains mostly of the original CP-odd state $A_{\rm CP-odd}$ (denoted as $A$ in the Lagrangian of Eq.~(\ref{eq:Aeff}) for simplicity)  and can be approximately expressed as:
\begin{equation}
    {A} \approx O_{A\pi^0}\pi^0+O_{A\eta}\eta+O_{A\eta'}\eta'+O_{AA}A_{\rm CP-odd}.
\label{eq:mesonmixing_1}
\end{equation} 
Here $O_{Ai}$ is the unitary transformation
matrix from gauge eigenstates to mass eigenstates. The expressions for $O_{Ai}$ are given in     Ref.~\cite{Domingo:2016yih}.   $O_{Ai}$ are typically small, except in the resonance region when $m_A\sim m_i$ for $i=\pi^0, \eta,$ and $\eta^\prime$.  This mixing effect contributes to additional production and decay channels of $A$, comparing to the case of the CP-even scalar $\phi$.

\subsection{Productions}
\label{sec:cp-odd-pro}

\begin{description}
\item[Production via Pseudoscalar Meson Mixing]    
Due to the mixing between $A_{\rm CP-odd}$ and pseudoscalar mesons of the SM,
any process that produces those meson states would also produce the new CP-odd scalar $A$. Following Ref.~\cite{FASER:2018eoc}, we can estimate its production cross section as
\begin{equation}
\label{eq:Apro}
\sigma_A \approx |O_{A\pi^0}|^2 \sigma_{\pi^0} +|O_{A\eta}|^2 \sigma_\eta+|O_{A\eta^\prime}|^2 \sigma_{\eta^\prime},
\end{equation}
where the values and distributions of cross sections $\sigma_{\pi^0}$, $\sigma_\eta$ and $\sigma_{\eta^\prime}$ are obtained from Ref.~\cite{Kling:2021fwx}.

\item[$B$ Meson and Kaon Decay]
The CP-odd scalar can also be produced in the decays of mesons, in particular $K \to \pi A$ and $B \to X_s A$~\cite{Dolan:2014ska}, similarly to the CP-even case, through effective flavor changing interactions.     We define the effective Lagrangian of flavor changing quark interactions with the CP-odd scalar $A$   as~\cite{Hall:1981bc, Frere:1981cc, Freytsis:2009ct, Batell:2009jf, Dolan:2014ska}   
\begin{equation}
    \mathcal{L}_{eff}
    =-i\frac{A}{v}\sum\xi^{ij}_{A}m_{f_j}\bar{f}_i P_R f_j+h.c..
    \label{eq:Lfceff_A}
\end{equation}
  The explicit form of $\xi_A^{ij}$ depends on how the CP-odd scalar embedded in the model.  The expression from the 2HDM contributions is given in Sec.~\ref{sec:2HDM_A}.   

With this effective interaction, the branching fractions of $B$-meson decaying into $K^{(*)}A$ are given by~\cite{Freytsis:2009ct}  
\begin{align}
    \text{Br}(B\rightarrow KA) 
    &=\frac{1}{\Gamma_B}\frac{G_F|\xi_A^{sb}|^2}{32\sqrt{2}\pi}\frac{(m_B^2-m_K^2)^2\left[f_0(m_{A}^2)\right]^2}{m_B^3}\left[\lambda(m_B^2, m_K^2, m_A^2)\right]^{1/2},   \label{eq:B_to_A}\\
    \text{Br}(B\rightarrow K^*A)& 
    =\frac{1}{\Gamma_B}\frac{G_F|\xi_A^{sb}|^2}{32\sqrt{2}\pi}\frac{\left[A_0(m_{A}^2)\right]^2}{m_B^3}\left[\lambda(m_B^2, m_{K^*}^2, m_A^2)\right]^{3/2},  
\end{align}
 where the function $\lambda(a,b,c)=(a-b-c)^2-4bc$ and the form factors $f_0$ and $A_0$ can be found in Ref.~\cite{Ali:1999mm}.  The branching fraction of the inclusive $B\to X_s A$ is given at leading order of $\Lambda_{\rm QCD}/m_b$ by~\cite{Freytsis:2009ct}
\begin{equation}
    \text{Br}(B\rightarrow X_s A) 
    =\frac{1}{\Gamma_B}\frac{G_F|\xi_A^{sb}|^2}{16\sqrt{2}\pi}m_b^3\left(1-\frac{m_A^2}{m_b^2}\right).
\end{equation}
The branching fractions of kaon decaying into $\pi A$ can be expressed similar to those in Eq.~(\ref{eq:B_to_A})~\cite{Dolan:2014ska}.

\item[Radiative Bottomonium $\Upsilon$ and Charmonium $J/ \psi$ decays]
 
 A light pseudoscalar can be produced in the radiative decay of bottomonium $\Upsilon\rightarrow\gamma A$, or charmonium $J/ \psi \rightarrow\gamma A$. It is convenient to express the corresponding branching ratio in the form of~\cite{Winkler:2018qyg,BESIII:2021ges,Domingo:2010am}  
\begin{equation}
    \frac{\text{Br}({\Upsilon\rightarrow\gamma A})}{\text{Br}({\Upsilon\rightarrow \ell^+\ell^-})}=\frac{ G_F m_b^2|{\xi_A^b}|^2}{\sqrt{2}\pi\alpha_{\rm ew}}(1-\frac{m_A^2}{m_\Upsilon^2})\times C_{\text{QCD}}^b,
\end{equation}
and similarly for $\text{Br}({J/\psi\rightarrow\gamma A})$.
Here  $C_{\text{QCD}}^{b}$ includes the QCD correction to the leptonic width of ${\Upsilon\rightarrow \ell^+\ell^-}$, as well as $m_A$ dependent QCD and relativistic corrections to the decay of ${\Upsilon\rightarrow\gamma A}$~\cite{Drewes:2015iva,Beneke:1997jm,McKeen:2008gd}.  
\end{description}

\subsection{Decays of light CP-odd Scalar}
\label{sec:Adecay}

 We list below the dominant decay channels for $A$ in different $m_A$ region.    
For $m_A<1.3$ GeV, the interaction of CP-odd scalar $A$ with pseudo-Goldstone bosons can be derived using chiral perturbation theory~\cite{Domingo:2016yih}.  For 1.3 GeV$<m_A<3$ GeV, the spectator model is employed with  partonic dynamics while keeping the kinematics of hadrons.  For $m_A>3$ GeV, we use the spectator model at parton level to find the decay width into quark or gluon pairs.
 
\begin{description}
 
\item[Decays into Diphoton:]
 
Given that the mass eigenstate $A$ is a mixture of the CP-odd scalar $A_{\rm CP-odd}$ and pseudo-goldstone bosons $\pi^0$, $\eta$ and $\eta^\prime$, the contribution to $A\rightarrow \gamma\gamma$ includes the contribution induced from the mixing as shown in Eq.~(\ref{eq:mesonmixing_1}). The effective couplings analogous to $\xi_A^\gamma$ in Eq.~(\ref{eq:xiAgamma}) but for $\pi^0$, $\eta$ and $\eta'$ obtained from experiments are
\begin{equation}
    C_{A}^\gamma=\xi_{A}^\gamma/v,\ \ 
    C_{\pi^0}^\gamma=-10.75\ \rm{GeV}^{-1},\  \ 
    C_{\eta}^\gamma=-10.8\ \rm{GeV}^{-1},\ \ 
    C_{\eta'}^\gamma=-13.6\ \rm{GeV}^{-1}.
\end{equation}
The decay width of $A\to \gamma\gamma$ is given by 
\begin{equation}
    \Gamma(A\to \gamma\gamma)=\frac{\alpha_{\rm ew}^2m_A^3}{64\pi^3}\biggl|O_{AA}C_A^\gamma+O_{A\pi^0}C_{\pi^0}^\gamma+O_{A\eta}C_{\eta}^\gamma+O_{A\eta'}C_{\eta'}^\gamma\biggr|^2.
\end{equation}

\item[Decays into Leptons]

The leptonic decay width of $A$ is given by
\begin{equation}
    \Gamma(A\to \ell^+\ell^-)=\frac{G_F m_A m_\ell^2 \beta_\ell}{4\sqrt{2}\pi } |\xi_A^{\ell}|^2,
\end{equation}
with $\ell=e, \mu, \tau$ and $\beta_\ell=\sqrt{1-4m_\ell^2/m_A^2}$, since contributions from meson mixing are small enough to be neglected.

\item[Hadronic Decays into Tri-meson for $\mathbf{m_A \lesssim 1.3}~\gev$:] 
The decay width for a pseudoscalar $A$ to tri-meson final state $\Pi_i\Pi_j\Pi_k$   may be written as
\begin{equation}
\begin{aligned}
    \Gamma(A\to \Pi_i\Pi_j\Pi_k)&=\frac{1}{256S_{ijk}\pi^3m_A}\int_{(m_j+m_k)^2}^{(m_A-m_i)^2}ds  |\mathcal{M}_A^{ijk}|^2\\ &\sqrt{1-\frac{2(m_j^2+m_k^2)}{s}+\frac{(m_j^2-m_k^2)^2}{s^2}}\times\sqrt{\biggl(1+\frac{s-m_i^2}{m_A^2}\biggr)^2-\frac{4s}{m_A^2}},
\end{aligned}
\label{eq:trimeson}
\end{equation}
where  $S_{ijk}$ is a symmetry factor: 1, 2, $3!$ depending on the number of identical particles in the final state. $\mathcal{M}_A^{ijk}$ stands for the transition amplitude for process $A\to \Pi_i\Pi_j\Pi_k$.  Note that since the mass eigenstate $A$ is a mixture of $\pi^0$, $\eta$, $\eta^\prime$ and CP-odd state $A_{\rm CP-odd}$ as shown in Eq.~(\ref{eq:mesonmixing_1}), $\mathcal{M}_A^{ijk}$ receives contribution not only from $A_{\rm CP-odd}\to \Pi_i\Pi_j\Pi_k$,  denoted as $\mathcal{A}_A^{ijk}$, 
but also from quartic-meson transition amplitude  $\mathcal{A}^{ijkl}$:
 \begin{equation}
\mathcal{M}_A^{ijk}\propto  O_{AA}\mathcal{A}_A^{ijk} + \sum_l O_{Al}\mathcal{A}^{ijkl}.
\end{equation}
Expressions for $\mathcal{A}_A^{ijk}$ are collected in Appendix~\ref{app:cpoddAdecay} while $\mathcal{A}^{ijkl}$ can be directly calculated from standard chiral perturbation theory, which can be found in Ref.~\cite{Domingo:2016yih}.

\item[Radiative Hadronic Decays for $\mathbf{m_A \lesssim 1.3}~\gev$:]
The radiative decay of $A\to \pi^+\pi^-\gamma$ at leading order are introduced by the mixing effect as shown in Eq.~(\ref{eq:mesonmixing_1}) as well and  can not  be neglected.   The  $\pi^+\pi^-\gamma$ partial decay width of pseudoscalar $A$ is given by 
\begin{equation}
    \Gamma(A\to \pi^+\pi^-\gamma)=\int_{4m_\pi^2}^{m_A^2}ds \Gamma_0(s) |O_{A\eta}B_\eta(s)+O_{A\eta'}B_{\eta'}(s)|^2.
\end{equation}
Expressions for $\Gamma_0(s)$, $B_\eta(s)$, and $B_{\eta^\prime}(s)$ can be found in Refs.~\cite{Holstein:2001bt, M.A.B.:1972, Venugopal:1998fq}, with all $m_{\eta/\eta'}$ replaced by $m_A$.   This radiative decay could be important for  $m_A\sim m_{\eta, \eta^\prime}$.

\item[Hadronic Decays for $\mathbf{1.3\  GeV \lesssim m_A \lesssim 3}~\gev$ (Spectator Model):] 

At $m_A>1.3$ GeV, the decay widths predicted by chiral perturbation theory become less reliable. As a transition to the perturbative partonic decay, for $1.3$ GeV  $\lesssim m_A \lesssim 3$ GeV, we adopt  spectator model with  partonic dynamics while keeping the kinematics of the hadrons~\cite{McKeen:2008gd,Dolan:2014ska,Holstein:2001bt}.

The effective Lagrangian for the interactions of $A$ with the partons in the spectator model is 
\begin{equation}
    \mathcal{L}_{\rm spect.}=\frac{i}{\sqrt{2}} A_1(\mathcal{Y}_u^A\bar u\gamma_5 u+\mathcal{Y}_d^A\bar d\gamma_5 d+\mathcal{Y}_s^A\bar s\gamma_5 s), 
\end{equation}
with
\begin{equation}
    \mathcal{Y}_u^A\approx \frac{\sqrt{2}B }{\sqrt{3}vf_\pi^2}m_u\xi_A^u,\quad \mathcal{Y}_d^A\approx \frac{\sqrt{2} B}{\sqrt{3}vf_\pi^2}m_d\xi_A^d,\quad \mathcal{Y}_s^A\approx \frac{\sqrt{2}B }{\sqrt{3}vf_\pi^2}m_s\xi_A^s,
\end{equation}  
with $B(m_u+m_d)/(2f_\pi)=m_\pi^2\simeq (135{\ \rm MeV})^2$, $Bm_s/f_\pi= (m_{K^0}^2+m_{K^\pm}^2-m_\pi^2)\simeq (688{\ \rm MeV})^2$,  and $f_\pi \approx 93 $ MeV. We still use Eq.~(\ref{eq:trimeson}) to calculate the tri-meson decay width, with the decay amplitude $\mathcal{M}_A^{ijk}$ expressed using $\mathcal{Y}_{u,d,s}^A$ above, as shown in Ref.~\cite{Domingo:2016yih}.   

\item[Decays into Quarks for $m_A>3$ GeV]
We use the partonic decay widths into quarks and gluons for hadronic decays at higher pseudoscalar masses.  The ratios of the decay rates to quarks comparing to that to dilepton are given by  
\be
    \Gamma_{\bar{\ell}\ell}:\Gamma_{\bar{s}s}:\Gamma_{\bar{c}c}:\Gamma_{\bar{b}b}=
    (\xi_A^{\ell})^2  m_\ell^2\beta_\ell:3(\xi_A^s)^2 m_s^2\beta_s:3 (\xi_A^c)^2 m_c^2\beta_c:3 (\xi_A^b)^2 m_b^2\beta_b.
\ee

\item[Decays into Gluons for $m_A>3$ GeV]

Using the effective $Agg$ coupling defined in Eq.~(\ref{eq:Aeff}), the decay width of $A\to gg$ can be expressed as
\begin{equation}
    \Gamma(A\to gg)=\frac{G_F\alpha_s^2m_A^3}{4\sqrt{2}\pi^3}|\xi_A^g|^2.
\end{equation}
   
 \end{description}

 \section{Case Study: Type-I Two Higgs Doublet Model}
 \label{sec:2HDM}

%*****************************************************
\subsection{Model and Couplings}
%*****************************************************

Now we consider 2HDM as a case study, in which one of the neutral  non-SM scalars is very light.   The Higgs sector of the 2HDM~\cite{Branco:2011iw} consists of two SU(2)$_L$ scalar doublets  $\Phi_i\,(i=1,2)$ with hyper-charge $Y=1/2$
\begin{equation}
        \Phi_i = \left(\begin{array}{c}
                \phi_i^+\\
                (v_i+\phi_i^0+iG_i^0)/\sqrt{2}
        \end{array}\right),
\end{equation}
where $v_i\,(i=1,2)$ are the vacuum expectation values (vev) of the doublets after the electroweak symmetry breaking (EWSB), satisfying $v_1^2+v_2^2=v^2=(246\ {\rm GeV})^2$.

The Higgs potential in the Higgs sector of general CP-conserving 2HDM is
\begin{eqnarray}
 V(\Phi_1, \Phi_2) &=& m_{11}^2\Phi_1^\dag \Phi_1 + m_{22}^2\Phi_2^\dag \Phi_2 -m_{12}^2(\Phi_1^\dag \Phi_2+ h.c.) + \frac{\lambda_1}{2}(\Phi_1^\dag \Phi_1)^2 + \frac{\lambda_2}{2}(\Phi_2^\dag \Phi_2)^2  \notag \\
 & &+ \lambda_3(\Phi_1^\dag \Phi_1)(\Phi_2^\dag \Phi_2)+\lambda_4(\Phi_1^\dag \Phi_2)(\Phi_2^\dag \Phi_1)+\frac{1}{2}   \Big[\lambda_5 (\Phi_1^\dag \Phi_2)^2 + h.c.\Big]\,,
 \label{eq:higgspotential}
\end{eqnarray}
which is responsible for the EWSB, the Higgs masses, and the trilinear and quartic Higgs couplings. 
After the EWSB, the scalar sector of the 2HDM consists of five physical scalars:  two CP-even scalars $h$ and $H$, one CP-odd scalar $A$, and a pair of the charged ones $H^\pm$. In our discussion below, we take $h$ to be the SM-like Higgs.   It is convenient to replace the model parameters $(m_{11}^2, m_{22}^2,  \lambda_{1,2,3,4,5})$ by the physical Higgs masses, EWSB vev $v$, ratio of the Higgs vevs $\tan\beta=v_2/v_1$, and CP-even Higgs mixing angle $\alpha$: $(m_h, m_H, m_A, m_{H^\pm}, v, \tan\beta, \cos(\beta-\alpha))$. There is an additional soft $Z_2$ symmetry breaking term $m_{12}^2$, which is usually replaced by the parameter $\lambda v^2\equiv m_H^2-\frac{m_{12}^2}{\sin\beta\cos\beta}$   that enters the theoretical constraints.

Given different arrangements of $\Phi_1$ and $\Phi_2$ couplings to the SM quarks and leptons, there are four possibilities for 2HDMs: Type-I, Type-II, Type-L and Type-F~\cite{Branco:2011iw}.
For the Type-II, -L, and -F, each of the Higgs doublets couples to at least one type of quarks or leptons.  As a consequence, over the entire region of $\tan\beta$, there are always unsuppressed couplings of the scalars with at least one type of fermions.  Therefore, it is difficult to realize very weakly coupled long-lived scalars.    Thus in our study, we focus on the Type-I 2HDM, where only one Higgs doublet couples to all quarks and leptons.   All the fermion couplings are suppressed at large $\tan\beta$.  Specifically,  the  normalized couplings of the fermions with various Higgses at leading order are
\begin{eqnarray}
    \xi_h^f&=&\frac{\cos\alpha}{\sin\beta}=\sin(\beta-\alpha)+\cosba \cot\beta, \nonumber \\
    \xi_H^f&=&\frac{\sin\alpha}{\sin\beta}=\cosba-\sin(\beta-\alpha) \cot\beta, \nonumber \\ 
    \xi_A^f&=&\cot\beta {\rm \ for\ }f=u, \ \ \ -\cot\beta {\rm\ for\ }f=d,e. 
\end{eqnarray}
The loop induced $\xi_{H}^{g, \gamma}$ couplings of the non-SM CP-even Higgs $H$ depend on $m_H$.   The full expressions are given at Eq.~(\ref{eq:coup_hgg}) for contributions to $\xi_{H}^{g}$ from quarks  and Eq.~(\ref{eq:coup_hgaga}) for contributions $\xi_{H}^{\gamma}$ from charged quarks/leptons and $W$.     In the 2HDM, there are additional contributions to $\xi_{H}^{\gamma}$ from charged Higgses with coupling term of $\lambda_{HH^+H^-}H H^+H^-$:
\begin{equation}
    \xi_H^\gamma|_{H^\pm} = -\frac{v \lambda_{H H^{+} H^{-}}}{2m_{H^{\pm}}^{2}} \mathcal{A}_{0}^{\phi}\left(\tau_{H^{\pm}}\right),
    \label{eq:coup_hgaga_Hpm}
\end{equation}
while there is no such contribution to $\xi_A^\gamma$ given the lack of $AH^+H^-$ coupling.

%*****************************************************
\subsection{Theoretical and Experimental Constraints}
%*****************************************************
 
 In this section, we consider various theoretical and experimental constraints on the Type-I 2HDM, and identify the regions of parameter space in which a light weakly coupled neutral scalar can be accommodated.
%*****************************************************

\subsubsection{Unitarity and Vacuum Stability}

We consider theoretical constraints of unitarity, perturbativity and vacuum stability. Detailed discussion of the theoretical constraints can be found in Ref.~\cite{Kling:2016opi,Chen:2019pkq}. Given the current LHC measurements of the SM-like Higgs couplings~\cite{Su:2019ibd}, as well as the requirements of long-lived light scalar as discussed below, a small $|\cosba|$ close to the alignment limit of $\cos(\beta-\alpha)\sim0$ is necessary. 

 Vacuum stability   sets a lower bound on $\lambda v^2\equiv m_H^2-\frac{m_{12}^2}{\sin\beta\cos\beta} \gtrsim 0$ as well as the lower limit on the mass splitting $m_{H^\pm/A}^2-m_H^2$~\cite{Kling:2016opi}. The unitarity and perturbativity together set the upper bounds on variables such as the mass splitting of $m_{H^\pm/A}^2-m_H^2$, $\lambda v^2$ and $\tan\beta$:  
\be
\lambda v^2&< 4\pi v^2,\\
\max\{\tan\beta,\cot\beta\} &\lesssim \sqrt{(8\pi v^2)/(3\lambda v^2)},\\
m_{H^\pm/A}^2-m_H^2 &\lesssim \mathcal{O}\left(4\pi v^2-\lambda v^2\right).
\label{eq:unitarity}
\ee
The allowed range for $\tan\beta$ is strictly bounded for large $\lambda v^2$ and unbounded when $\lambda v^2=0$.  Given that the couplings of $H/A$ to fermions are  proportional to $1/\tan\beta$,  $\lambda v^2\sim 0$ is preferred for a weakly coupled light Higgs to have a suppressed couplings to fermions.  In addition, the lower and upper bounds for $m_{H^\pm/A}-m_H^2$ are determined solely by $\lambda v^2$.   
  
\begin{figure}[htb]
\centering
\includegraphics[clip,trim=4mm 0cm 4mm 0mm,width=0.45\textwidth]{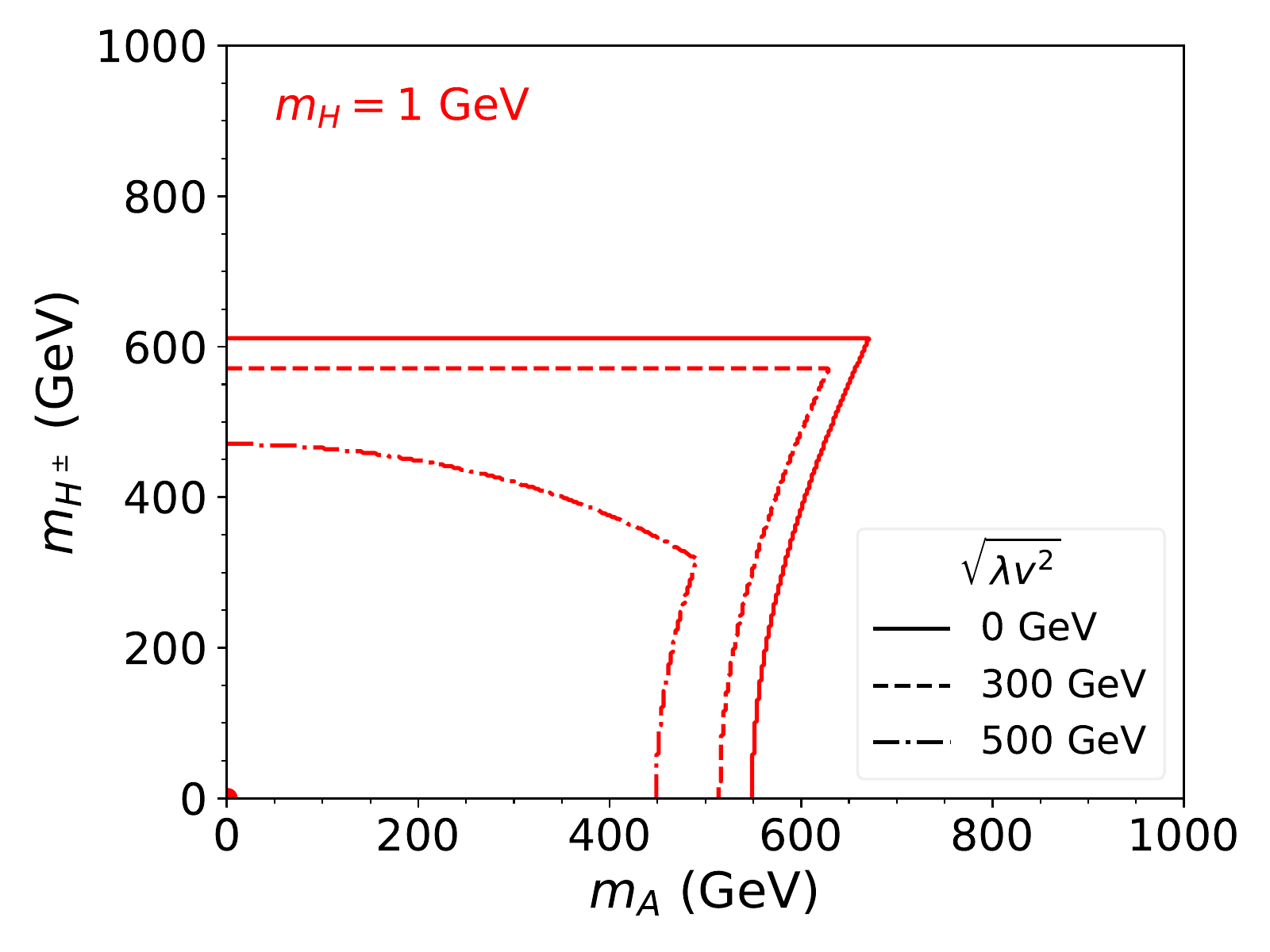} 
\includegraphics[clip,trim=4mm 0cm 4mm 0mm,width=0.45\textwidth]{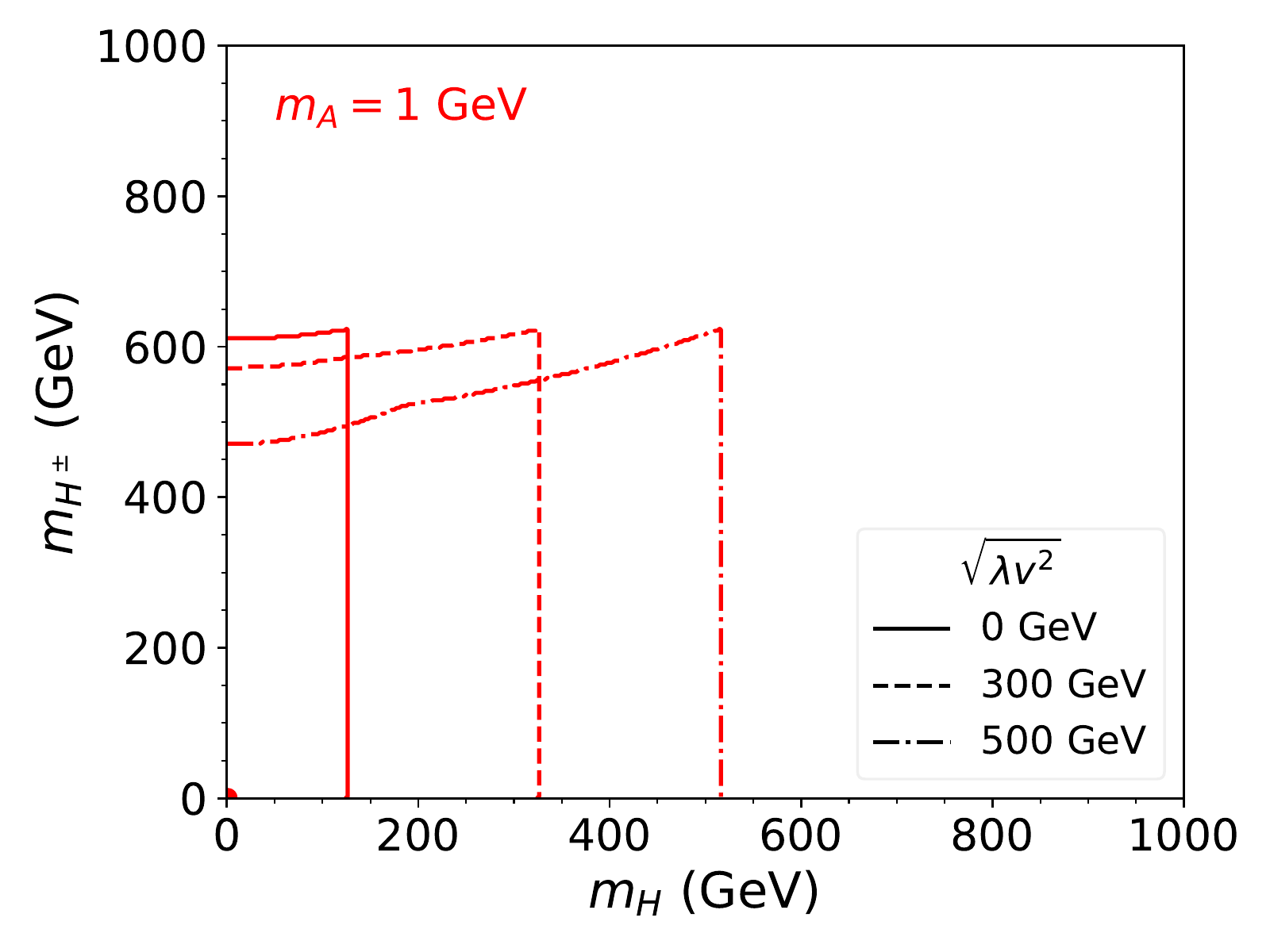} 
\caption{Allowed region below and to the left of the curves by theoretical constraints for $m_H=1$ GeV (left panel) and $m_A=1$ GeV (right panel) for various values of $\lambda v^2$. Here we have $\cosba=0$.}
\label{fig:theoretical_constraint}
\end{figure}

To explore the scenarios in which a light non-SM Higgs is allowed, in Fig.~\ref{fig:theoretical_constraint}, we plot the allowed region (below and to the left of the curves) in the plane of $m_{H^\pm}$ and $m_{A/H}$ for $m_H=1$ GeV (left panel) and $m_A=1$ GeV (right panel) under the alignment limit of $\cos(\beta-\alpha)=0$.  In order for $H$ to be light, i.e. $m_{H}\sim 1$ GeV, the heavy Higgs mass $m_{H^\pm/A}$ can not be higher than around 600 GeV, and the maximally allowed charged Higgs mass is achieved when $\lambda v^2=0$. Note that the allowed regions are not very sensitive to $m_H$ for small $m_H$, so the conclusion holds for any $m_H$ around zero. In the right panel for $m_A=1$ GeV,  $m_H$ is restricted to be less than 125 GeV at $\lambda v^2=0$ while $m_{H^\pm}$ is allowed to reach around 600 GeV.  Thus we conclude that, by the considerations of the theoretical constraints, the weakly coupled light neutral scalar is only allowed in two scenarios with $\lambda v^2\approx 0$:
\begin{eqnarray}
m_H\sim 0: && m_{A/H^\pm}\lesssim 600\ {\rm GeV}  \\
m_A\sim 0: && m_{H^\pm}\lesssim 600\ {\rm GeV},  \quad m_H\lesssim m_h.
\end{eqnarray}
The conclusion holds for small $|\cosba|\sim 0$ as well.

%*****************************************************
\subsubsection{Electroweak Precision Constraints}

 The current precisions on  the oblique parameters $S$, $T$, $U$ as well as the correlations among them are \cite{Haller:2018nnx}
\be
&S=0.04 \pm 0.11,\quad && T=0.09\pm 0.14,\quad &&U=-0.02 \pm 0.11, \\
&\rho_{ST}=0.92, && \rho_{SU}=-0.68,&& \rho_{TU}=-0.87.
\ee
The electroweak precision measurements impose strong constraints on the mass splittings between the neutral and charged scalars of the Higgs doublet:  $m_{H^\pm}$ need to be around the mass of either $m_H$ or $m_A$~\cite{Gu:2017ckc,Chen:2018shg,Cheng:1989ib}. 
In particular, for small $m_H$, only $m_A \sim m_{H^\pm}$ is allowed. 

Combining with theoretical constraints and the direct searches at LEP~\cite{ALEPH:2013htx},  the legitimate scenarios for weakly coupled light scalars are~\footnote{Ref.~\cite{ALEPH:2013htx} only shows constraints for $m_A>12$ GeV. Roughly speaking, smaller $m_A$ leads to weaker constraints on $m_{H^\pm}$. }
\begin{eqnarray}
m_H\sim 0: && m_{A}\sim m_{H^\pm}\lesssim 600\ {\rm GeV},
\label{eq:light_mH}
\\
m_A\sim 0: &&   m_{H^\pm}\sim m_H\lesssim m_h,
\label{eq:light_mA}
\end{eqnarray}
with $\lambda v^2\approx 0$ and $|\cosba|\sim 0$.

%*****************************************************
\subsubsection{Flavor Constraints}

The flavor observations,  such as $B\rightarrow X_s \gamma$, $B_{s,d}\rightarrow \mu^+\mu^-$, $B-\bar{B}$ mixing, decays of $B$ and $D$ baryons, impose strong constraints on the charged Higgs mass as well as the value of $\tan\beta$.  The limits on charged Higgs mass for four types of 2HDMs have been thoroughly studied in Ref.~\cite{Haller:2018nnx}.  Unlike the Type-II and Type-F 2HDMs with a charged Higgs mass $m_{H^\pm}<800$ GeV   excluded by the measurement of the branching fraction of $B\to X_s\gamma$~\cite{Atkinson:2021eox,Misiak:2020vlo}, in the Type-I 2HDM, only the low $\tan\beta$ region receives flavor constraints.  The strongest bound comes from $B_d \rightarrow \mu^+\mu^-$, which excludes regions of $\tan\beta<3$ for charged Higgs mass of 100 GeV.  The constraints get weaker for larger $m_{H^\pm}$: $\tan\beta<1.2$  for $m_{H^\pm}=800$ GeV.

%---------------------------------------------------------
\subsubsection{Invisible Higgs Decays}
%---------------------------------------------------------

For a light $H/A$ with long lifetime, $h\to HH/AA$ is constrained from the
invisible Higgs decay of ${\rm Br}(h\to {\rm invisible})<0.24$~\cite{Khachatryan:2016whc,Aad:2015txa,Aaboud:2017bja}.   
The Branch fraction of Higgs invisible decay is given by~\cite{Feng:2017vli} 
\begin{eqnarray}
{\rm Br}(h\to HH/AA) &=& \frac{\Gamma(h\to HH/AA)}{\Gamma_h} \nonumber \\
&\approx&  \frac{1}{\Gamma_h^{\rm SM}}\frac{g_{hHH/hAA}^2}{8\pi m_h^2}\biggl(1-\frac{4m_{H/A}^2}{m_h^2}\biggr)^{1/2}\simeq 4700\cdot \biggl(\frac{g_{hHH/hAA}}{v} \biggr)^2.
\label{eq:brhphiphi}
\end{eqnarray}

The full expressions for $hHH$ and $hAA$ couplings can be found at Eqs.~(\ref{eq:ghHH}) and (\ref{eq:ghAA}). To achieve suppressed $g_{hHH}$ or $g_{hAA}$ to satisfy the invisible Higgs decay constraints,  we have
\begin{eqnarray}
    {\rm Light} ~H: &&\cosba=\tan 2\beta\frac{2\lambda v^2+m_h^2}{2(m_H^2-3\lambda v^2-m_h^2)} 
    \approx \frac{1}{\tan\beta}\, , \\
    {\rm Light} ~A: &&\cosba=\tan 2\beta\frac{2\lambda v^2+m_h^2+2m_A^2-2m_H^2}{2(m_H^2-\lambda v^2-m_h^2)} \approx \frac{1}{\tan\beta} \frac{2 m_H^2- m_h^2}{ m_H^2-m_h^2} \ ,
    \label{eq:BRh_zero}
\end{eqnarray}
at the leading order of $\cosba$, under the approximation of large $\tanb$, small $\lambda v^2$, and light $m_H$ or $m_A$.  

For the light $H$ under this limit, $g_{hHH}\approx -\frac{m_h^2}{4v} c_{\beta-\alpha}^2$, which leads to ${\rm Br}(h\to HH) \simeq 75 c_{\beta-\alpha}^4$. The experimental bounds on the invisible decay branching ratio of 0.24 can be satisfied for $c_{\beta-\alpha}<0.25$ and $\tan\beta>4$.  
At the same time, the couplings of $H$ to gauge bosons and fermions are suppressed as well for the large $\tan\beta$ region of the Type-I 2HDM:  
\begin{eqnarray}
&& \xi_{A}^{f}=1/\tanb,
 \label{eq:light_Acoup}
\\
&&\xi_{H}^{V}=c_{\beta-\alpha}\approx 1/\tanb,\\
&&\xi_{H}^{f}=c_{\beta-\alpha}(1- s_{\beta-\alpha}) \approx 1/(2 \tan^3 \beta).
 \label{eq:light_Hcoup}
\end{eqnarray}
Therefore, diphoton channel becomes dominated when $\tanb$ gets large at the Type-I 2HDM.  

For the light pseudoscalar scenario, we could adopt the same way as the light $H$ case discussed above to meet Higgs invisible decay constraint. However,  $g_{hAA}$ can also stay small under alignment limit when $A$ is light, and $m_H\sim m_h/\sqrt{2}\sim 90$ GeV.
Combining with other constraints that lead to ~\autoref{eq:light_mH} and ~\autoref{eq:light_mA}, we consider two benchmark scenarios in the Type-I 2HDM,
\begin{eqnarray}
    {\rm Light} ~H: &&\cosba=\frac{1}{\tan\beta}, \ m_A=\mC=600 \gev ,\ \lambda v^2=0 \, , 
    \label{eq:BM_H}\\
    {\rm Light} ~A: &&\cosba=0, \ m_H=m_{H^\pm}=90 \gev, \  \lambda v^2=0   \ ,
    \label{eq:BM_A}
\end{eqnarray}
with large $\tanb$ to accommodate a long lived particle.    Note that in the light $A$ case, a relatively light charged Higgs of 90 GeV is chosen.  Such a scenario with GeV-scale pseudoscalar survives the LEP charged Higgs search~\cite{ALEPH:2013htx}: $m_{H^\pm}\gtrsim 85$ GeV is still viable for light $m_A$.  The LHC charged Higgs search~\cite{ATLAS:2021xhq,ATLAS:2018gfm,CMS:2019bfg,CMS:2019idx,Cheung:2022ndq,Hu:2022gwd} only excluded $\tan\beta<5$ for $m_{H^\pm}$ in the mass range of (100,160) GeV.

 %*****************************************************

%*****************************************************
\subsubsection{Other Experimental Constraints}
%*****************************************************
\label{sec:constraints}
There are a variety of constraints on light scalars from beam dump experiments, supernovae, and meson decays.  Here we have a brief list summarizing the most relevant ones.  

\begin{description}

    \item [CHARM bounds] The CHARM Collaboration has searched for light axion-like particles at CERN with a 400 GeV proton beam-dump experiment on a copper target~\cite{CHARM:1985anb}. Its results can be used to constrain the light scalar~\cite{Winkler:2018qyg,Gorbunov:2021ccu}.
    \item [SuperNova] 
    A light, weakly coupled scalar can affect astrophysical processes. During supernova (SN) explosion, the scalar emission can  contribute significantly to the energy loss, shortening the neutrino pulse duration~\cite{Turner:1987by}.  Observation of core energy loss from the emission of light scalars produced through nucleon bremsstrahlung process  $N N \to N N S(A)$, would place constraints on the light scalars~\cite{Ellis:1987pk,Krnjaic:2015mbs,Batell:2019nwo}.   
    \item [$B$ meson decays] For (pseudo)scalar mass below the $B$ threshold, searches for  $B$ decays with leptonic final states become relevant. The leading constraints come from LHCb measurements of  $B \to K^* \phi $ with $\phi \to \mu\mu$ ~\cite{LHCb:2015nkv} and $B^{+} \rightarrow K^{+} \chi\left(\mu^{+} \mu^{-}\right)$~\cite{LHCb:2016awg}.  
    \item [Kaon  decays]  Kaon decays also contribute to the searches for light scalar region. The latest relevant ones are $K^{+} \rightarrow \pi^{+} X$ with $X$  to $\nu\bar\nu$ at   NA62~\cite{NA62:2021zjw} (90\% C.L.), $K^{+} \rightarrow \pi^{+} \chi\left(e^{+} e^{-}\right)$ at MicroBooNE~\cite{MicroBooNE:2021usw} (95\% C.L.), and ${\rm Br}\left(K^{+} \rightarrow \pi^{+} X\right)$ at E949~\cite{BNL-E949:2009dza} (90\% C.L.). All of them provide constraints based on the light scalar decay lifetime hypotheses.   
    \item [D meson decays] The current limits can be found in PDG~\cite{ParticleDataGroup:2022pth}, as well as the recent LHCb results~\cite{Aaij:2020wyk}. Those are typically not included in light scalar constraints since in most models, Br$(D^+ \to \pi^+ \phi$) (corresponding to Br$(c \to u \phi)$) is rather small.  
   \item [LEP]   {\tt OPAL, ALEPH} and {\tt L3} searches on $e^+e^- \to Z^* \phi$ at the LEP   detected $ 3 \times 10^6$ hadronic $Z$ decays~\cite{L3:1996ome,ALEPH:1993sjl,OPAL:2007qwz}, which included both the prompt and  invisible/long-lifetime $\phi$ cases. When $m_{\phi} \le 2m_\mu$, $\phi$ with high momentum can escape the LEP detector to be an invisibly decaying scalar. For $m_{\phi} > 2m_\mu$, $\phi$ could decay promptly. Thus the LEP search results could constrain the light scalar scenario~\cite{Winkler:2018qyg,Clarke:2013aya}.  
\end{description}
To impose the experimental constraints mentioned above, we recast the existing bounds to the Type-I 2HDM parameter space for $B$, kaon, $D$ meson decays as well as the LEP search results. For the CHARM bounds and SuperNova constraints, we use the approximate results from the SM with an additional light scalar scenario~\cite{Winkler:2018qyg,Dobrich:2018jyi} since the detailed recast of these two bounds involves a complete analyses of all possible contributions in the framework of the Type-I 2HDMs, which is left for future study.

\subsection{FASER and FASER2}
FASER is a cylindrical detector with a radius of 10 cm and a length of 1.5 m, installed in tunnel TI12 located at 480 m away from the ATLAS IP~\cite{Feng:2017uoz, FASER:2018ceo, FASER:2018bac, FASER:2022hcn, FASER:2021ljd,FASER:2021cpr}.  It is designed to detect LLPs produced at the ATLAS IP,   traveling in the very forward region,  and decaying in FASER into two very energetic particles.  FASER has been taking data since summer, 2022.   During the Run 3 of the LHC,  it is expected to collect data from proton-proton collisions of about 150 ${\rm fb}^{-1}$ integrated luminosity.  Given the distinctive signature and low background environment,  FASER provides a unique opportunity to probe light particles with suppressed couplings~\cite{Feng:2017uoz, FASER:2018eoc, FASER:2018bac}. Unlike all the other proposed LLP experiments,  FASER is able to detect photons with a preshower detector placed in front of the FASER calorimeter~\cite{Feng:2018pew, Kling:2020mch}.  A high-resolution preshower upgrade is planned to be installed in next two years~\cite{Boyd:2803084}, which could further increase the sensitivity.  

At the HL-LHC with an integrated luminosity of 3 ${\rm fb}^{-1}$,   FASER will be upgraded to FASER 2 with a larger volume of the detector, potentially at the same location~\cite{FASER:2018eoc} or at FPF~\cite{Anchordoqui:2021ghd, Feng:2022inv} about 620 meters from the LHC IP.  FASER 2 will extended the reach of FASER by an order of magnitude or more.   

In our analyses below, we adopt the configuration of FASER 2 in the original proposal~\cite{FASER:2018eoc}, sitting  480 m away from the LHC IP: 
\begin{eqnarray}
&&\mathrm{FASER}: \Delta=1.5 \mathrm{~m}, \quad R=10 \mathrm{~cm}, \quad \mathcal{L}=150 ~\mathrm{fb}^{-1}, 
\label{eq:FASER}
\\
&&\mathrm{FASER ~2}: \Delta=5 \mathrm{~m}, \quad R=1 \mathrm{~m}, \quad \mathcal{L}=3 ~\mathrm{ab}^{-1}.
\label{eq:FASER2}
\end{eqnarray}
Here $\Delta$ and $R$ are the detector length and radius respectively.

%****************************************************
% %%%%%%%%%%%%%%%%%%%%%%%%%%%%%%%%%%%%%%%%%%%%%%
% %%%%%%%%%%%%%%%%%%%%%%%%%%%%%%%%%%%%%%%%%%%%%%
% %%%%%%%%%%%%%%%%%%%%%%%%%%%%%%%%%%%%%%%%%%%%%%
\subsection{Results for the Light CP-even Higgs in the Type-I 2HDM}

The productions of a light CP-even Higgs $H$ are mostly via the semileptonic decay of pions and kaons, or the hadronic decay of kaons, $\eta$, $B$ and $D$ mesons, as well as radiative decay of bottomonium $\Upsilon$ as discussed in Sec.~\ref{sec:cp-even-pro}.    
In our numerical analyses below, we only take into account the $\phi$  productions from $B$, kaon, and pion meson decays since the contribution from $D$ meson is eleven orders of magnitude smaller~\cite{Boiarska:2019jym}.

In the 2HDM, the effective flavor changing coupling $\xi_\phi^{ij}$ as defined in Eq.~(\ref{eq:Lfceff_phi})  is given by~\cite{Eilam:1989zm, Li:2014fea, Cheng:2015yfu, Arnan:2017lxi}    
 \begin{align}
    \xi_\phi^{ij}|_{\text{2HDM},h/H}&=-\frac{4G_F\sqrt{2}}{16\pi^2}\sum_k V_{ki}^* m_k^2\left[g_1(x_k, x_{H^\pm}) \begin{pmatrix}\sin(\beta-\alpha) \\ \cos(\beta-\alpha)\end{pmatrix} \right. \nonumber \\
    &\left. +g_2(x_k, x_{H^\pm}) \begin{pmatrix}\cos(\beta-\alpha)\\ -\sin(\beta-\alpha)\end{pmatrix}-g_0(x_k, x_{H^\pm}) \frac{2v}{m_W^2}\begin{pmatrix}\lambda_{h H^+ H^-} \\ \lambda_{H H^+ H^-}\end{pmatrix}\right] V_{kj}, \label{eq:xi_H_2HDM}
\end{align}
where the upper functions are for $h$ and lower ones are for $H$, and the trilinear couplings $\lambda_{(h,H)H^+ H^-}$ are defined in Appendix~\ref{Appx:trihiggs}, and the auxiliary functions $g_{0, 1, 2}$ in the Type-I 2HDM are given in Appendix~\ref{Appx:aux_func} with $x_k\equiv m_x^2/m_W^2$ and $x_H^\pm\equiv m_{H^\pm}^2/m_W^2$ where $m_x$ is the mass of the quark running in the loop.  

 There are also 2HDM charged Higgs contributions to the effective four-fermion-Higgs interaction similar to~\autoref{Fig:4Quark} and~\autoref{Eq:Lag_4FH}.
 In our calculation, we ignore such contributions since usually the couplings between charged scalar and first two generations of fermions are suppressed by the small values of the first two generation fermion masses.

% %%%%%%%%%%%%%%%%%%%%%%%%%%%%%%%%%

\begin{figure}[htb]
\begin{center}
\includegraphics[width=15 cm]{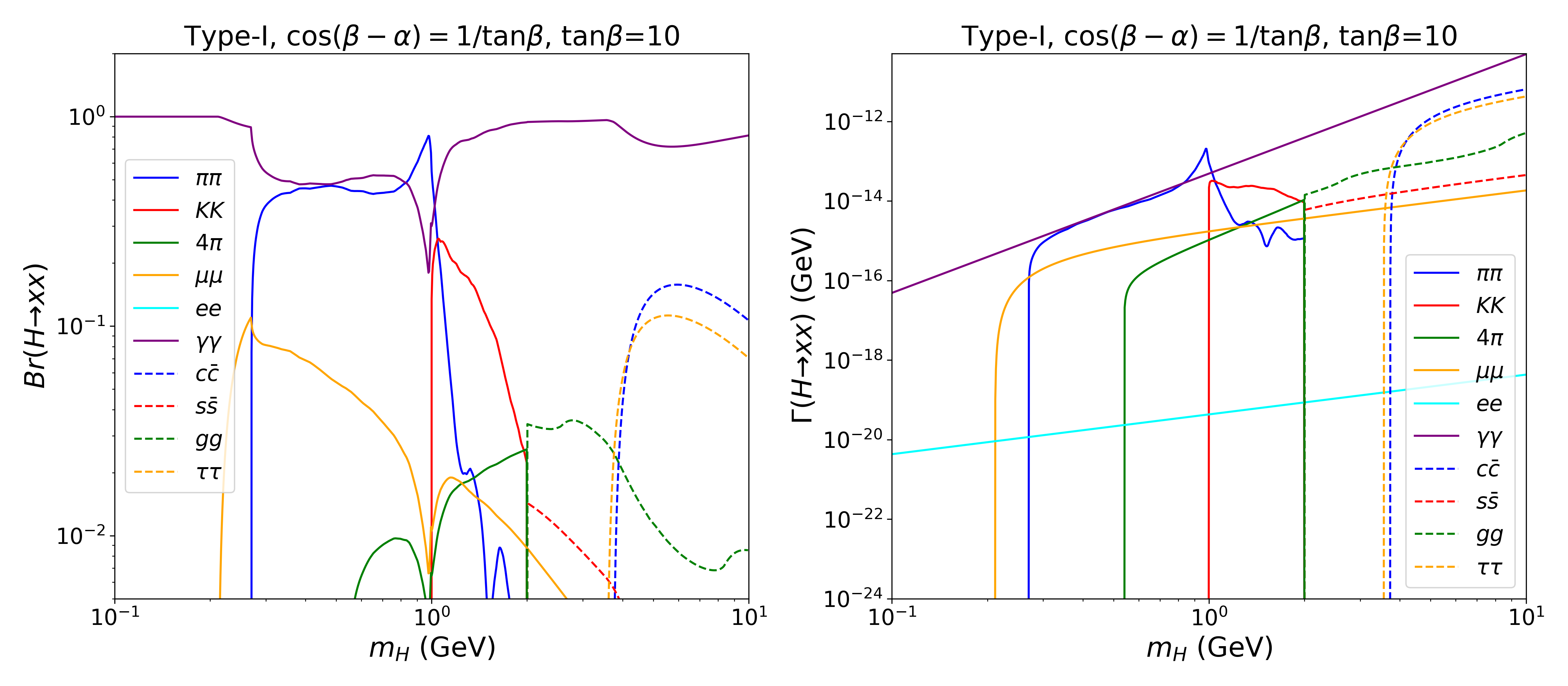}
\caption{The decay branching fractions (left) and partial widths (right) of light CP even Higgs  in the Type-I 2HDM  for the light $H$ benchmark point. Decays to hadrons and  quarks/gluons are connected at $m_H=2$ GeV.   }
\label{fig:TypeI_BR_cosba_tanb}
\end{center}
\end{figure}

Fig.~\ref{fig:TypeI_BR_cosba_tanb}  shows the decay branching fractions (left panel) and partial decay widths (right panel) of  the light $H$ in Type-I 2HDM under  the relation of $\cosba=1/\tanb$ for $\tanb=10$.
Here the dominant decay mode is diphoton, which receives $\tanb$ independent contributions from charged Higgs loop in addition.  All other channels into  the quark, lepton, and gluon final states are suppressed since $\xi_H^f \propto 1/\tan^3\beta$.  $H \to \pi\pi$ is dominated around 1 GeV due to the corresponding decay form factors~\cite{Winkler:2018qyg}. As discussed in Sec.~\ref{sec:Hdecay}, decays to mesons and quarks/gluons are connected smoothly at $m_H=2$ GeV.

\begin{figure}[htb]
\begin{center}
 \includegraphics[width=0.5\textwidth]{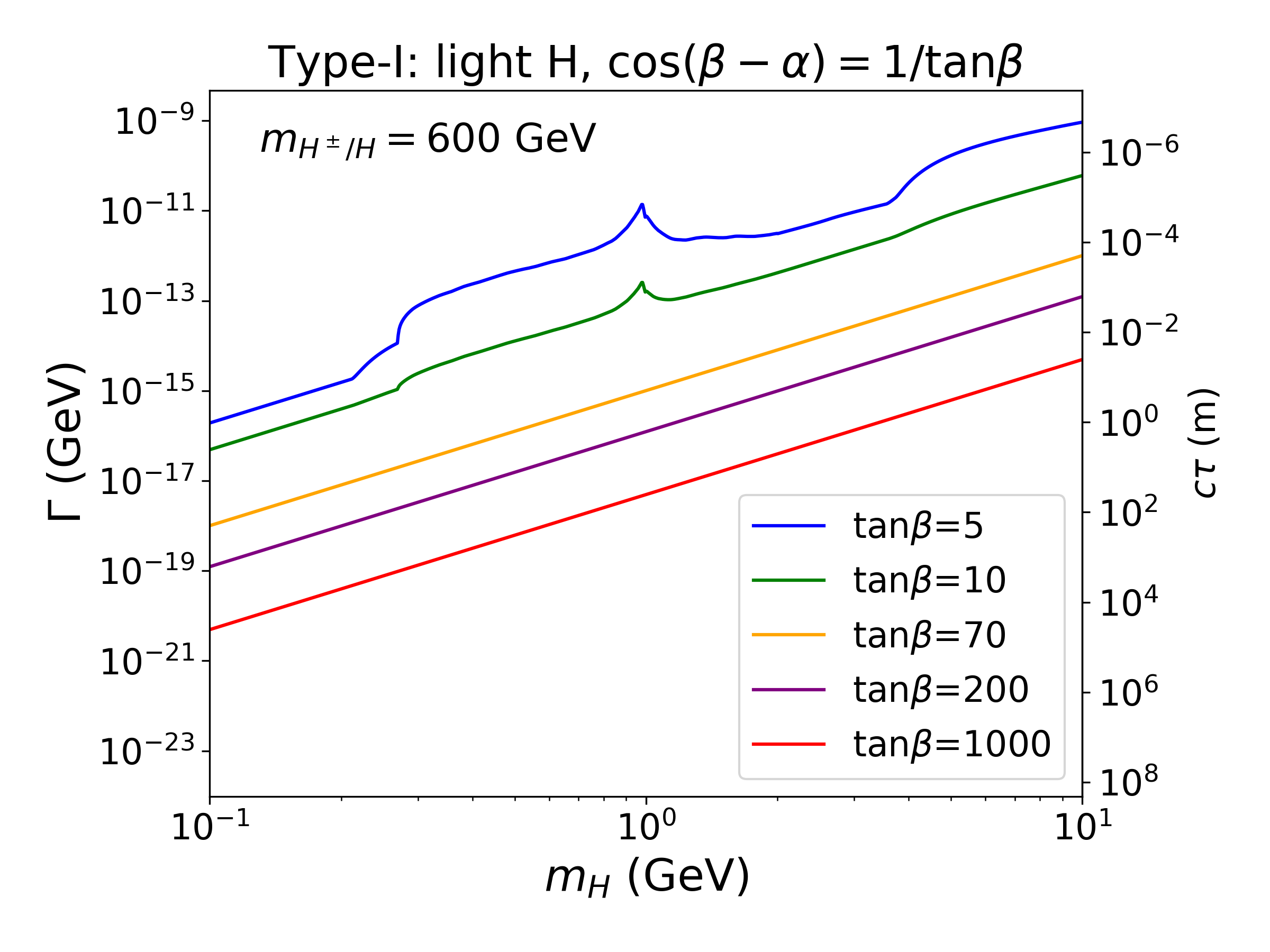}
\includegraphics[width=0.49\textwidth]{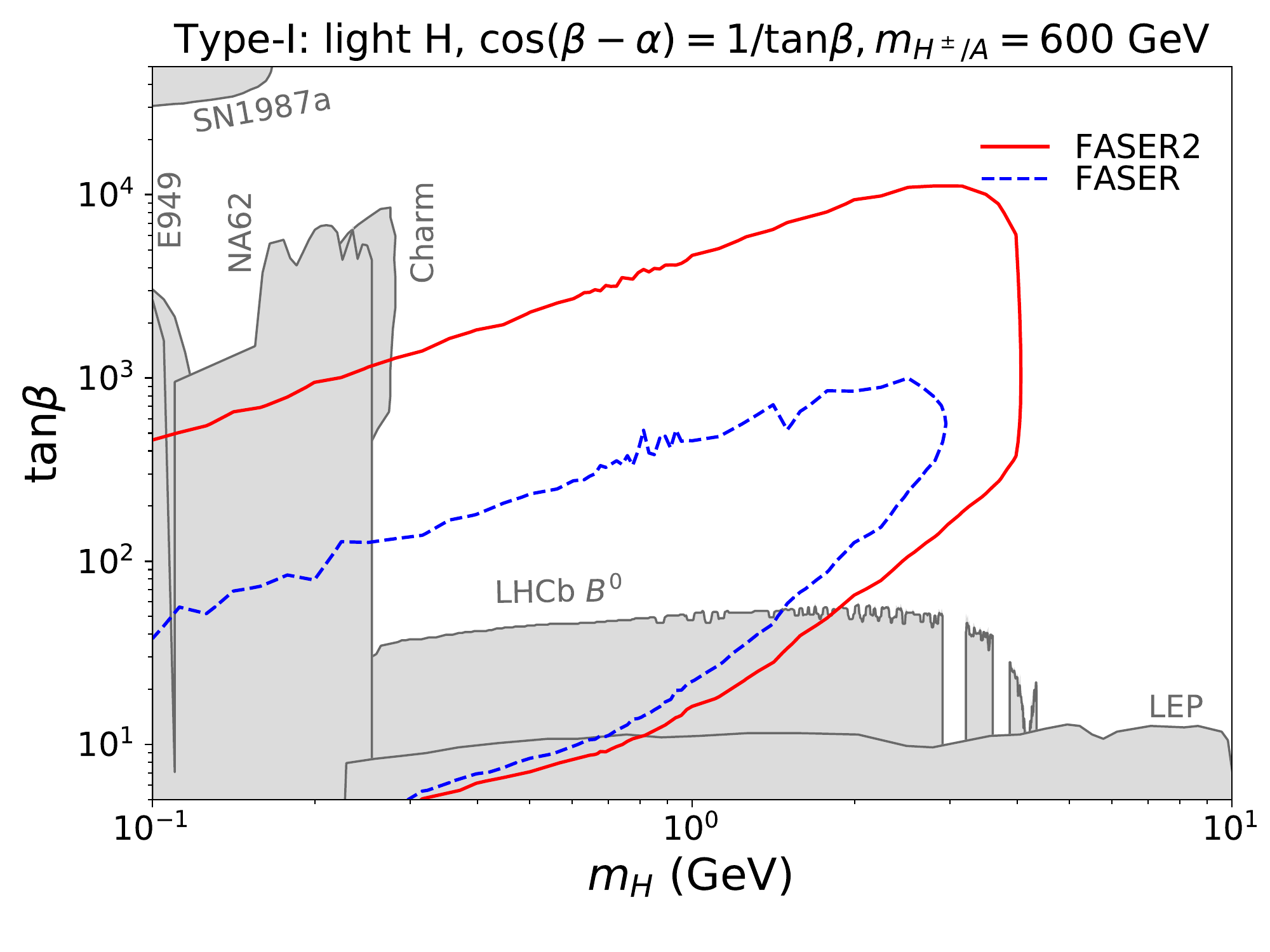}
 \caption{Left Panel: the total decay width (left $y$-axis) and decay length $c\tau$ (right $y$-axis) of the light CP-even Higgs  in the Type-I 2HDM for  the light $H$ benchmark point.
Right Panel:   FASER (blue dashed curve) and FASER 2 reach (red solid curve)  for the light CP even Higgs $H$ in the  $m_{H}$ vs. $\tan\beta$ plane. Various current experimental constrains are shown in grey regions.
\label{fig:TypeI_Gamma_cosba_tanb}
}
\end{center}
\end{figure}

The left panel of \autoref{fig:TypeI_Gamma_cosba_tanb} shows the decay width and  decay length $c\tau$ of the light $H$ in the Type-I 2HDM for the light $H$ benchmark point.  The $\Gamma$ and $c\tau$ become straight line for very large $\tan\beta$ as a consequence of dominated diphoton decay. $c\tau$ reaches a few centimeters to meters for $\tanb > 10$.  

To obtain the FASER and FASER 2  reaches, we consider the LLPs produced from the various meson decays with {\tt FORESEE}~\cite{Kling:2021fwx}. The light meson spectra are generated by {\tt EPOS-LHC}~\cite{Pierog:2013ria} as implemented in the package CRMC~\cite{crmc}, while the $B$ meson spectrum is generated by Pythia 8~\cite{Sjostrand:2007gs}. We assume 100\% acceptance rate for all final states with the FASER and FASER 2 configurations and integrated luminosities specified in Eqs.~(\ref{eq:FASER}) and (\ref{eq:FASER2}).
 
In the right panel of \autoref{fig:TypeI_Gamma_cosba_tanb}, we show the potential three event reach by FASER (blue dashed curve) and FASER2 (red solid curve) in the plane of  $m_H$ vs. $\tanb$ for the light $H$ benchmark point.   Also shown in gray regions  are the other   experimental constraints, including  $B$ meson decays at LHCb~\cite{LHCb:2015nkv}, $K^+$ decays from NA62~\cite{NA62:2021zjw},  and E949~\cite{BNL-E949:2009dza} (at 90\% C.L.), CHARM beam dump~\cite{CHARM:1985anb,Krnjaic:2015mbs}, light scalar search $e^+e^- \to Z^* \phi$ at LEP($L3$)~\cite{L3:1996ome,Winkler:2018qyg}, and light scalar constraints from supernova explosion at SN1987a~\cite{Turner:1987by,Ellis:1987pk,Krnjaic:2015mbs}.
The dip in the NA62 bounds around $m_\pi$ is  due to the cross over of two experimental search regions. 
MicroBooNE~\cite{MicroBooNE:2021usw} bounds, SM Higgs coupling measurements~\cite{Han:2020lta}, Higgs invisible decay, as well as flavor bounds~\cite{Haller:2018nnx} do not constrain the CP-even scalar case in the chosen parameter region of $\tan\beta>5$.  
 
While the sub-GeV region is already well explored by other current experiments, as well as the low $\tanb$ region by LHCb and LEP, FASER and FASER 2 offer unique opportunities to cover the large $\tan\beta$ region up to $\tanb=10^3$ and $10^4$ respectively, and $m_H$ reach up to $m_B$ due to the $B\to H X_s$ production.  FASER 2 increases the FASER reach in $\tan\beta$ by about one order of magnitude at large $\tanb$ region. Such a difference mainly comes from the 20 times luminosity enhancement in FASER 2.     The improvement due to the  larger detector size of FASER 2  mainly shows up at the large $m_H$ region, which pushes the limit to $m_H\approx m_B$.

 %------------------------
\subsection{Results for the Light CP-odd Higgs in the Type-I 2HDM}
\label{sec:2HDM_A}
%------------------------

Given the mixture  of the light CP-odd scalar with pseudo-Goldstone bosons $\pi^0$, $\eta$ and $\eta^\prime$ as shown in~\autoref{eq:Apro}, $A$ can be produced in any process that produces those mesons.    In addition,  $A$ can be produced in the weak decays of SM mesons, in particular $K \rightarrow \pi A$ and $B \rightarrow X_s A$, as well as the less important radiative decays of bottomonium $\Upsilon$ and charmonium $J/\psi$.  For the detailed formulae about the production, see Sec.~\ref{sec:cp-odd-pro}. In our numerical analyses, we take into account all of these productions except for the   radiative decays. 

Similar to the  CP-even  Type-I 2HDM case, the effective flavor changing  coupling $\xi_A^{ij}$ from \autoref{eq:Lfceff_A} is given by~\cite{Hall:1981bc, Frere:1981cc, Freytsis:2009ct, Eilam:1989zm, Li:2014fea, Cheng:2015yfu, Arnan:2017lxi} 
\begin{equation}
    \xi_A^{ij}|_{\rm 2HDM}=\frac{4\sqrt{2}G_F}{16\pi^2}\sum_k V^*_{ki}m_k^2\left[Y_1\left(x_k, x_{H^\pm}\right)\cot\beta+Y_2\left(x_k, x_{H^\pm}\right)\cot^3\beta\right]V_{kj}, \label{eq:xi_A_2HDM}
\end{equation}
for $ij$ being down-type quarks. The auxiliary functions $Y_{1, 2}\left(x_k, x_{H^\pm}\right)$ are given in Appendix~\ref{Appx:aux_func}.   

\begin{figure}[htb]
\begin{center}
    \includegraphics[width=1\textwidth]{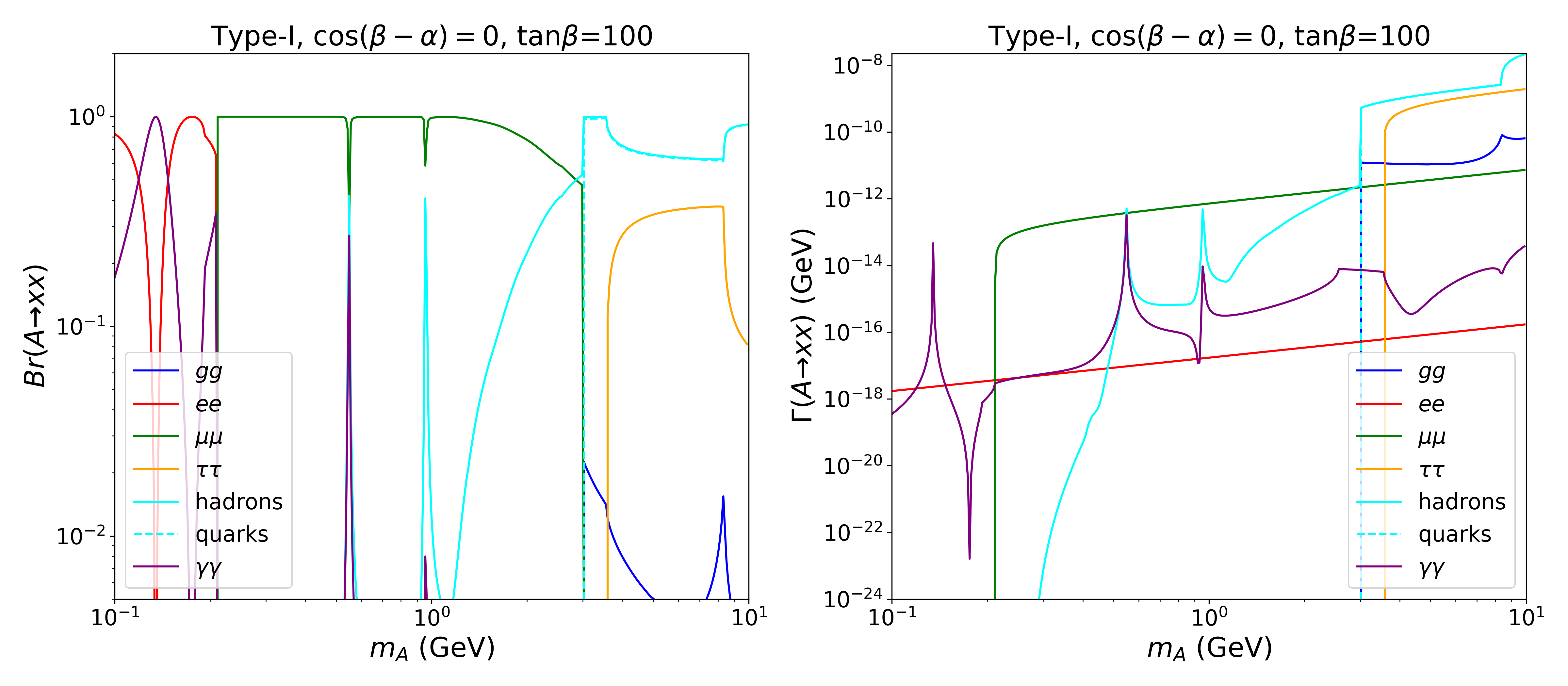}
\caption{The decay branching fractions (left panel) and partial decay widths of the light CP-odd Higgs $A$ in the Type-I 2HDM  for the light $A$ benchmark point with $\tanb=100$.   Decays to hadrons and quarks/gluons are connected at $m_A=3$ GeV.   }
\label{fig:TypeI_A_cosba_tanb}
\end{center}
\end{figure}

\autoref{fig:TypeI_A_cosba_tanb}  shows the decay branching fractions (left panel) and partial decay widths of the light CP-odd $A$ in the Type-I 2HDM for the light $A$ benchmark point with $\tanb=100$. For $m_A < 2\m_\mu$,   both $ee$ and $\gamma\gamma$ channels are important. $\mu\mu$ channel is dominated before hadronic modes open.  Once $m_A > 3$ GeV, hadronic decay modes dominate.

\begin{figure}[htb]
\centering
\includegraphics[width=0.5\textwidth]{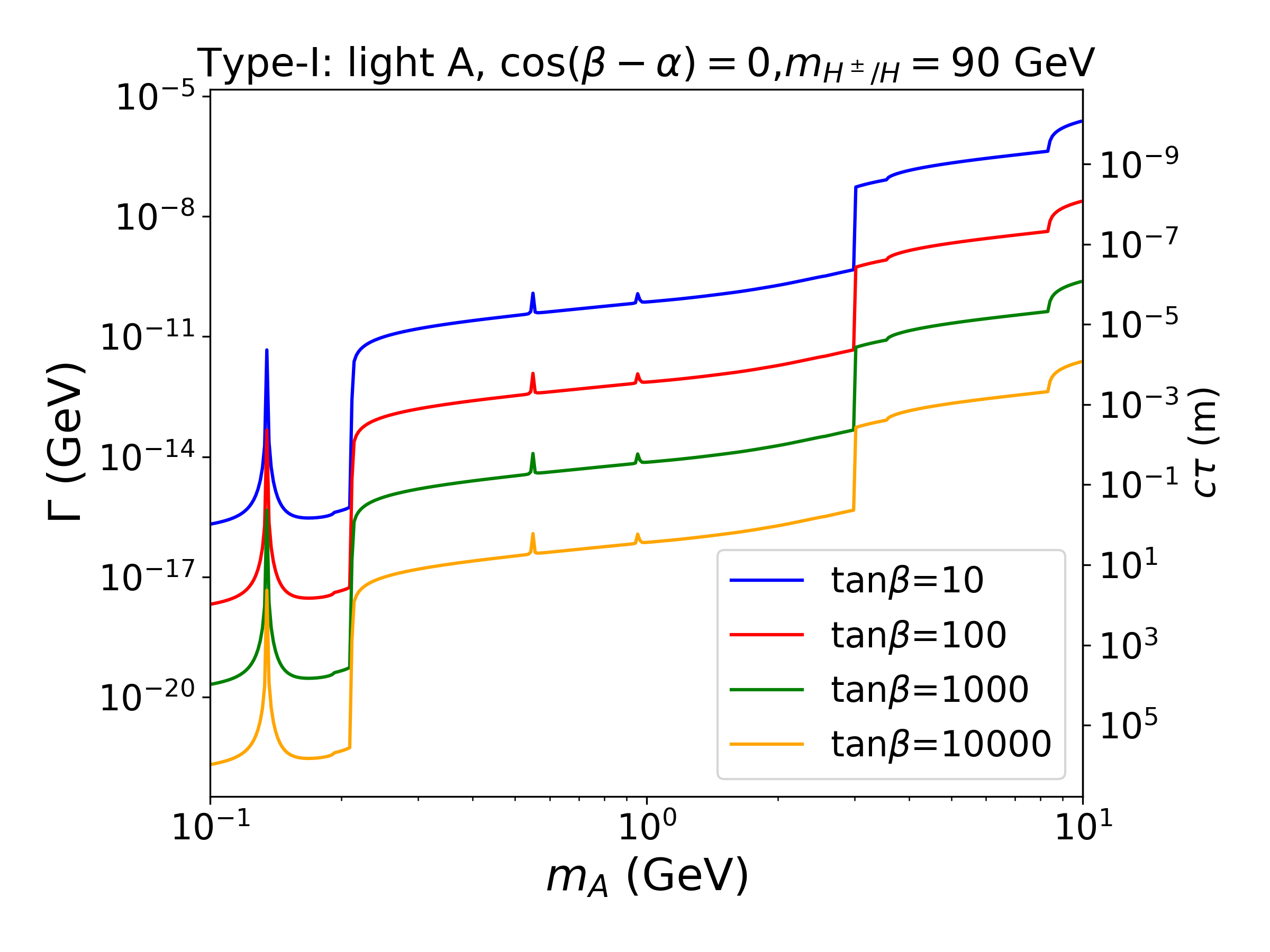}
\includegraphics[width=0.49\textwidth]{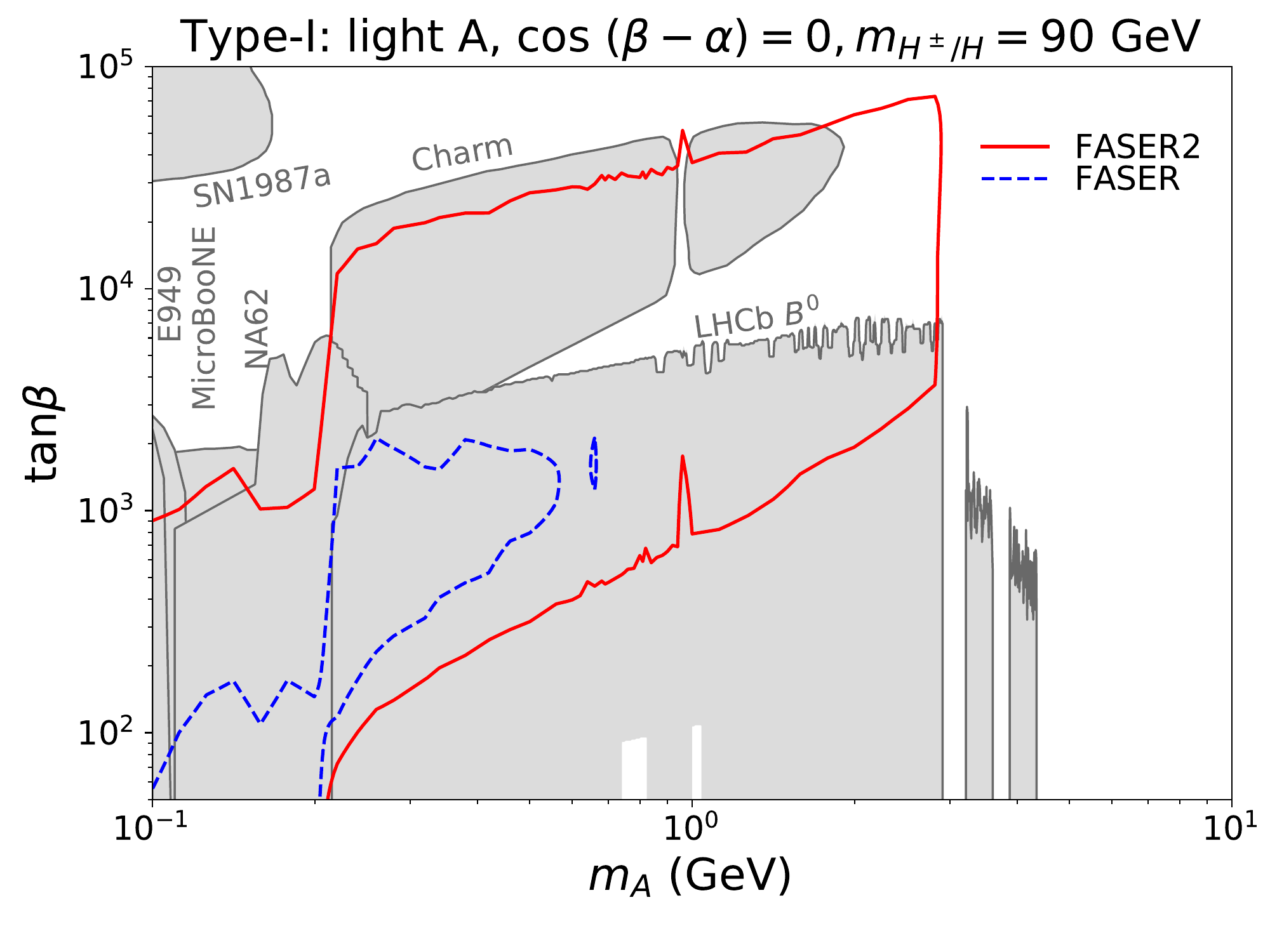}
\caption{Left Panel: the total decay width (left $y$-axis) and decay length $c\tau$ (right $y$-axis) of the light CP-odd Higgs  in the Type-I 2HDM for  the light $A$ benchmark point.
Right Panel: The FASER (blue dahsed curve) and FASER 2 reach (red solid curve)  for the light CP-odd Higgs $A$ in the parameter space of $m_{A}$ vs. $\tan\beta$ plane. Various current experimental constrains are shown in grey regions. 
}
\label{fig:lightA_deacay_lifetime}
\end{figure}

The decay  width  and decay  length $c\tau$ of the light pseudoscalar $A$ in  the Type-I 2HDM for various $\tan\beta$ are presented in the left panel of \autoref{fig:lightA_deacay_lifetime}. The peaks around $m_A\sim$ 1 GeV or below are introduced by the $\pi^0$, $\eta$ and $\eta^\prime$ resonances.  
The sudden increase of hadronic decay width at $m_A=3$ GeV is due to the transition from spectator model to perturbative theory of partons. After that point, the $c\bar c$ and gluon-gluon decays kick in, which leads to the  growth of total decay width~\footnote{The partonic approximation of the hardonic decay width at $m_A>3$ GeV is not very accurate. As pointed out in \cite{Domingo:2010am}, at the $c\bar c$ and $b\bar b$ thresholds, the CP-odd Higgs mixes with charmonium and bottomonium pseudoscalar states, which need to be accounted. Also, the kinematics at these threshold are three-body decay to $D\bar D\pi$ or $B\bar B \pi$ rather than a two-body decay.}.    Note  that unlike the CP-even case as shown in the left panel of Fig.~\ref{fig:TypeI_Gamma_cosba_tanb}, the $\tan\beta$ dependence of decay width $\Gamma_A$ is only of an overall shift with the same feature. This is because the couplings of $A$ to the SM particle has the identical  $1/\tan\beta$ dependence.

 In the right panel of \autoref{fig:lightA_deacay_lifetime}, we show the potential reach by FASER (blue dashed curve) and FASER 2 (red solid curve) in the plane of $m_A$ vs. $\tanb$ for the light $A$ benchmark point.  The other current experimental constraints are shown in gray regions, similar to Fig.~\ref{fig:TypeI_Gamma_cosba_tanb}.  Note that the CHARM bounds extends to the larger region of $m_A\sim 2$ GeV~\cite{Dobrich:2018jyi}, comparing to the CP-even case of  $m_H\sim 300$ MeV~\cite{Winkler:2018qyg}.    The LEP limits at the low $\tan\beta$  are  not present since the value of $\tan\beta$ starts at 50.   Comparing to the light $H$ case, regions with much larger $\tan\beta$ can be probed.  This is because the different $\tan\beta$ dependence for $\xi_A^f$, comparing to that of $\xi_H^f$, as shown in Eqs.~(\ref{eq:light_Acoup}) and (\ref{eq:light_Hcoup}).  Similarly to light CP-even scalar case, the FASER 2 coverage of $\tan\beta$ is about one order {of magnitude higher. The $m_A$ coverage in the light CP-odd scalar case is much more sensitive to the geometry of detector, especially its radius.  FASER only reaches $m_A$ about 0.7 GeV, while FASER 2 could reach till the production threshold of $m_B$.

\section{Conclusion}
\label{sec:conclusion} 

In this paper, we studied the scenario with a light weakly coupled CP-even scalar $\phi$ or a CP-odd scalar $A$ with a relatively long lifetime. We considered a model-independent framework describing the most general interactions between a CP-even or CP-odd scalar and SM particles using the notation of coupling modifiers in the effective Lagrangian.  We developed a general formalism for  the productions of the light scalar from meson decays, as well as  re-analysed the scalar decay rates.   In particular, we performed state of the art calculation of the hadronic decays of light scalars across different mass range, using chiral perturbation theory, dispersive analysis, and spectator model.  We also developed a general program~\cite{LSDedcay} to calculate the decays of a light CP-even or CP-odd scalar, incorporating the coupling modifiers of the light scalars to the SM particles.  Our program can be used to evaluate the decay of light scalars in many  scenarios beyond the SM. 

After developing the general formalism, we carried out a specific case study in the large $\tan\beta$ region of the Type-I 2HDM, which could naturally accommodate a light scalar with suppressed couplings while satisfying all the theoretical and experimental constraints. We chose two benchmark scenarios: a light $H$ with $\cosba=1/\tanb$ and other non-SM scalar mass around 600 GeV, and a light $A$  under alignment limit $\cosba=0$ and other non-SM scalar mass 90 GeV.  The light scalar decay length varies in $(10^{-8}, 10^5)$ meters. We further  obtained the FASER and FASER 2 reaches for those benchmark scenarios, which probe the parameter space at the very large $\tan\beta$ region.  The comparison of the FASER and FASER 2 reach shows that higher luminosity helps to  reach the weaker coupling region. A larger detector, especially the radius helps to extend the reach in $m_A$. The current FASER 2 configuration can reach the mass production threshold around $m_B$.

Forward LHC experiments, like FASER and FASER 2, offer a unique opportunity to detect light long-lived sectors.    They are complementary to the beyond the SM  searches based on  the prompt decay at the LHC main detectors, LLP searches in the transverse region, as well as fixed target searches at low energies.  The discovery of a long-lived light scalar at FASER and FASER 2 provides an unambiguous evidence for new physics beyond the SM. The on-going LHC Run 3 and the upcoming HL-LHC have great potential in exploring both the energy and intensity frontiers of particle physics.

\begin{acknowledgments}

%%%%%%%%%%%%%%%%%%%%%%%%%%%%%%%%%%%%%%%%%%%%%%%%%%%%%%%%%%%%%%%%%%%%%%%%%%%%%%%%%%%%%%
We would like to thank Felix Kling for his participation in this project at the early stage and insightful discussions later on. HS is supported by the International Postdoctoral Exchange Fellowship Program. 
SS is supported  by the Department of Energy under Grant No.~DE-FG02-13ER41976/DE-SC0009913.
WS is supported by the Junior Foundation of Sun Yat-sen University and Shenzhen Science and Technology Program (Grant No. 202206193000001, 20220816094256002).  

\end{acknowledgments}

%++++++++++++++++++++++++++++++++++++++++++++++++++++++++
% \clearpage
\appendix 

\section{Form Factors $A_{0,1/2,1}$}
\label{app:phigg_phigaga}
The expressions for the form factors $A_{0,1/2,1}^\phi$ for scalars, fermions, and gauge bosons in the loop contribution to $\xi_\phi^g$ and $\xi_\phi^\gamma$ of the CP-even scalar $\phi$ are~\footnote{Note that this definition of $A_{0,1/2,1}^\phi$ differs from Ref.~\cite{Djouadi:2005gi} by a factor of 1/2.}
\begin{eqnarray}
&&A_{0}^{^\phi}(\tau) =-\frac{1}{2}[\tau-f(\tau)]\tau^{-2}, \\
&&A_{1 / 2}^{^\phi}(\tau) =[\tau+(\tau-1) f(\tau)] \tau^{-2}, \\
&&A_{1}^{^\phi}(\tau) =-\frac{1}{2}\left[2 \tau^{2}+3 \tau+3(2 \tau-1) f(\tau)\right] \tau^{-2}, 
\label{eq:Aeven}
\end{eqnarray}
with $\tau=m_{\phi}^{2} / 4 m^{2}$ for $m$ being the mass of the particle running in the loop, and 
\begin{equation}
f(\tau)=\left\{\begin{array}{ll}
\arcsin ^{2} \sqrt{\tau} & \text { if } \tau \leqslant 1 \\
-\frac{1}{4}\left(\log \frac{1+\sqrt{1-1 / \tau}}{1-\sqrt{1-1 / \tau}}-i \pi\right)^{2} & \text { if } \tau>1
\end{array}\right.  .
\end{equation}

For the CP-odd scalar $A$, the form factor $A_{1/2}^A$ for fermion loop contribution to $\xi_A^g$ and $\xi_A^\gamma$ of the CP-odd scalar $A$ is 
\begin{equation}
\mathcal{A}_{1/2}^A=2\tau^{-1}f(\tau).
\label{eq:Aodd}
\end{equation}

\section{Formulae Related to Tri-meson Decay of CP-odd Scalar}
\label{app:cpoddAdecay}

The decay width for a pseudoscalar $A$ to tri-meson final state $\Pi_i\Pi_j\Pi_k$  can be written as
\begin{equation}
\begin{aligned}
    \Gamma(A\to \Pi_i\Pi_j\Pi_k)=&\frac{1}{256S_{ijk}\pi^3m_A}\int_{(m_j+m_k)^2}^{(m_A-m_i)^2}ds  |\mathcal{M}_A^{ijk}|^2\\ &\sqrt{1-\frac{2(m_j^2+m_k^2)}{s}+\frac{(m_j^2-m_k^2)^2}{s^2}}\times\sqrt{\biggl(1+\frac{s-m_i^2}{m_A^2}\biggr)^2-\frac{4s}{m_A^2}},
\end{aligned}
\end{equation}
where $m_{A,i,j,k}$ are the masses for $A$, $\Pi_i$, $\Pi_j$, $\Pi_k$, respectively. $S_{ijk}$ is a symmetry factor: 1, 2, $3!$ depending on the number of identical particles in the final state. $\mathcal{M}_A^{ijk}$ stands for the transition amplitude for process $A\to \Pi_i\Pi_j\Pi_k$,  which receives contributions   from $A_{\rm CP-odd}\to \Pi_i\Pi_j\Pi_k$, denoted as $\mathcal{A}_A^{ijk}$, as well as from quartic-meson transition amplitude  $\mathcal{A}^{ijkl}$ due to mixing.   
\begin{equation}
\mathcal{M}_A^{ijk}\propto  O_{AA}\mathcal{A}_A^{ijk} + \sum_l O_{Al}\mathcal{A}^{ijkl}.
\end{equation}

Expressions for $\mathcal{A}_A^{ijk}$   can be read off from the chiral Lagrangian~\cite{Domingo:2016yih}~\footnote{ Ref.~\cite{Domingo:2016yih} had a minus sign ahead of Eq.~(\ref{eq:A_Api0K0Kb0}) which should not be there.  }: 
\begin{eqnarray}
\mathcal{A}_A^{\pi^0\pi^0\pi^0}&=&3\mathcal{A}_A^{\pi^0\pi^+\pi^-}=-\frac{1}{2vf_\pi}\biggl(\frac{Bm_u}{f_\pi}\xi_A^u-\frac{Bm_d}{f_\pi}\xi_A^d\biggr),\\
\mathcal{A}_A^{\eta\pi^0\pi^0}&=&\mathcal{A}_A^{\eta\pi^+\pi^-}=-\frac{\sqrt{3} }{6vf_\pi}\biggl(\frac{Bm_u}{f_\pi}\xi_A^u+\frac{Bm_d}{f_\pi}\xi_A^d\biggr)(\cos\theta_\eta-\sqrt{2}\sin\theta_\eta),\\
\mathcal{A}_A^{\eta'\pi^0\pi^0}&=&\mathcal{A}_A^{\eta'\pi^+\pi^-}=-\frac{\sqrt{3} }{6vf_\pi}\biggl(\frac{Bm_u}{f_\pi}\xi_A^u+\frac{Bm_d}{f_\pi}\xi_A^d\biggr)(\sin\theta_\eta+\sqrt{2}\cos\theta_\eta),\\
\mathcal{A}_A^{\pi^0\eta\eta}&=&-\frac{1}{6vf_\pi}\biggl(\frac{Bm_u}{f_\pi}\xi_A^u-\frac{Bm_d}{f_\pi}\xi_A^d\biggr)(\cos\theta_\eta-\sqrt{2}\sin\theta_\eta)^2,\\
\mathcal{A}_A^{\pi^0\eta'\eta'}&=&-\frac{1}{6vf_\pi}\biggl(\frac{Bm_u}{f_\pi}\xi_A^u-\frac{Bm_d}{f_\pi}\xi_A^d\biggr)(\sin\theta_\eta+\sqrt{2}\cos\theta_\eta)^2,\\
\mathcal{A}_A^{\pi^0\eta\eta'}&=&-\frac{1}{6vf_\pi}\biggl(\frac{Bm_u}{f_\pi}\xi_A^u-\frac{Bm_d}{f_\pi}\xi_A^d\biggr)(\cos\theta_\eta-\sqrt{2}\sin\theta_\eta)(\sin\theta_\eta+\sqrt{2}\cos\theta_\eta),\\
\mathcal{A}_A^{\pi^0 K^+K^-}&=&-\frac{1}{6vf_\pi}\biggl(\frac{2Bm_u}{f_\pi}\xi_A^u+\frac{Bm_s}{f_\pi}\xi_A^s\biggr),\\
\mathcal{A}_A^{\pi^0 K^0\bar K^0}&=&\frac{1}{6vf_\pi}\biggl(\frac{2Bm_d}{f_\pi}\xi_A^d+\frac{Bm_s}{f_\pi}\xi_A^s\biggr), \label{eq:A_Api0K0Kb0}\\
\mathcal{A}_A^{\pi^+ K^- K^0}&=&\mathcal{A}_A^{\pi^- K^+ \bar K^0}=-\frac{\sqrt{2}}{6vf_\pi}\biggl(\frac{Bm_u}{f_\pi}\xi_A^u+\frac{Bm_d}{f_\pi}\xi_A^d+\frac{Bm_s}{f_\pi}\xi_A^s\biggr),
\end{eqnarray}
with $B(m_u+m_d)/(2f_\pi)=m_\pi^2\simeq (135{\ \rm MeV})^2$, $Bm_s/f_\pi= (m_{K^0}^2+m_{K^\pm}^2-m_\pi^2)\simeq (688{\ \rm MeV})^2$,  and $f_\pi \approx 93 $ MeV.

Expression for $\mathcal{A}^{ijkl}$ can be found in Ref.~\cite{Domingo:2016yih}.

%++++++++++++++++++++++++++++++++++++++++++++++++++++++++

\section{Tri-Higgs Couplings\label{Appx:trihiggs}}
  The trilinear couplings of $hHH$ and $hAA$ are: 
\begin{eqnarray}
g_{hHH}  &= &\frac{s_{\beta-\alpha}}{2v}\biggl[
(m_{H}^2 - 3\lambda v^2-m_h^2) ( 2 t_{2\beta}^{-1} s_{\beta-\alpha}c_{\beta-\alpha}  - c_{\beta-\alpha}^2 +s^2_{\beta-\alpha}) 
 + (\lambda v^2-m_H^2 ) 
\biggr],\label{eq:ghHH}\\
g_{hAA} 
&=&\frac{1}{2v}\biggl[(2 m_H^2 - 2 \lambda v^2 -2m_{A}^2-m_h^2)s_{\beta-\alpha}+2(m_H^2 - \lambda v^2-m_h^2)t_{2\beta}^{-1} c_{\beta-\alpha}\biggr] \ ,
\label{eq:ghAA}
\end{eqnarray}
with $\lambda v^2\equiv m_H^2-m_{12}^2/\cos\beta\sin\beta$.
The couplings of $h$ and $H$ to the charged Higgses are:
\begin{equation}
\lambda_{H^{+} H^{-} h} =\frac{1}{v}\left[\left(2 m_H^{2}-2 \lambda v^2 - 2 m_{H^{\pm}}^{2}-m_{h}^{2}\right) s_{\beta-\alpha}+2\left(m_H^{2}-\lambda v^2 - m_{h}^{2}\right) \cot 2 \beta c_{\beta-\alpha}\right], 
\end{equation}
\begin{equation}
\lambda_{H^{+} H^{-} H}=-\frac{1}{v}\left[-2 \lambda v^2  \cot 2 \beta s_{\beta-\alpha}+\left(2 m_{H^{\pm}}^{2}-m_{H}^{2}+2\lambda v^2 \right) c_{\beta-\alpha}\right].
\end{equation}
For $\lambda v^2=0$, we have
\begin{equation}
\begin{aligned}
\lambda_{H^{+} H^{-} H}&=-\frac{1}{v}\left(2 m_{H^{\pm}}^{2}-m_{H}^{2}\right) c_{\beta-\alpha}.
\label{eq:coup_HCC}
\end{aligned}
\end{equation}

\section{Auxiliary Functions for the 2HDM\label{Appx:aux_func}}
The auxiliary functions $g_{0, 1, 2}$ used in the effective flavor changing coupling $\xi_\phi^{ij}$ of the CP-even scalar defined in Eq.~(\ref{eq:xi_H_2HDM}) can be found in Ref.~\cite{Arnan:2017lxi} for different types of 2HDMs. For the Type-I 2HDM, they are reduced to
\begin{align}
    &g_0(x_k, x_{H^\pm})=-\cot^2\beta\frac{3x_{H^\pm}^2-4x_{H^\pm}x_k+x_k^2-2x_k(2x_{H^\pm}-x_k)\log\frac{x_{H^\pm}}{x_k}}{16 (x_{H^\pm}-x_k)^3}, \\
    &g_1(x_k, x_{H^\pm})=-\frac{3}{4}+\cot^2\beta\frac{x_k\left[5x_{H^\pm}^2-8x_{H^\pm}x_k+3x_k^2-2x_{H^\pm}(2x_{H^\pm}-x_k) \log\frac{x_{H^\pm}}{x_k}\right]}{4(x_{H^\pm}-x_k)^3}, \\
    &g_2(x_k, x_{H^\pm})=\cot\beta X_1(x_k, x_{H^\pm})+\cot^3\beta X_2(x_k, x_{H^\pm}),
\end{align}
with
\begin{align}
    X_1(x_k, x_{H^\pm})=&-\frac{1}{4}\left\{ \left[\frac{x_{H^\pm}}{x_{H^\pm}-x_k}-\frac{6}{(x_k-1)^2}+3\right]-\frac{x_{H^\pm}(3x_{H^\pm}-2 x_k)}{(x_{H^\pm}-x_k)^2}\log x_{H^\pm} \right. \nonumber \\
    &\left. +\left[\frac{x_{H^\pm}(3x_{H^\pm}-2x_k)}{(x_{H^\pm}-x_k)^2}+\frac{3 (x_k+1)}{(x_k-1)^3}\right]\log x_k\right\}, \\
    X_2(x_k, x_{H^\pm})=&\frac{x_k(5x_{H^\pm}-3x_k)}{4(x_{H^\pm}-x_k)^2}-\frac{x_{H^\pm}x_k(2 x_{H^\pm}-x_k)}{2 (x_{H^\pm}-x_k)^3}\log\frac{x_{H^\pm}}{x_k}.
\end{align}

For the CP-odd case, the auxiliary functions $Y_{1, 2}$ used in the effective flavor changing coupling $\xi_A^{ij}$  (Eq.~\ref{eq:xi_A_2HDM}) are
\begin{align}
    Y_1\left(x_k, x_{H^\pm}\right)=&\frac{1}{4}\left[-\frac{3x_{H^\pm}x_k-6x_{H^\pm}-2x_k^2+5x_k}{\left(x_k-1\right)\left(x_{H^\pm}-x_k\right)}+\frac{x_{H^\pm}\left(x_{H^\pm}^2-7x_{H^\pm}+6x_k\right)}{\left(x_{H^\pm}-x_k\right)^2\left(x_{H^\pm}-1\right)}\log x_{H^\pm}\right. \nonumber \\
    &\left.-\frac{x_{H^\pm}^2\left(x_k^2-2x_k+4\right)+3x_k^2\left(2x_k^2-2x_{H^\pm}-1\right)}{\left(x_{H^\pm}-x_k\right)^2\left(x_k-1\right)^2}\log x_k\right], \\
    Y_2\left(x_k, x_{H^\pm}\right)=&\frac{1}{2}\left[\frac{x_k}{x_{H^\pm}-x_t}-\frac{x_{H^\pm}x_k}{\left(x_{H^\pm}-x_k\right)^2}\log\frac{x_{H^\pm}}{x_t}\right].
\end{align}

\bibliography{FASER_light}

\end{document}